\documentclass[12pt]{article}
\catcode`@=11
\@addtoreset{equation}{section}
\renewcommand{\theequation}{\arabic{section}.\arabic{equation}}

\usepackage{amsmath}
\usepackage{amssymb}
\usepackage{amsthm}
\usepackage{psfrag}
\usepackage{graphicx}
\usepackage{color}
\usepackage{upgreek}
\DeclareMathAlphabet{\mathpzc}{OT1}{pzc}{m}{it}

\allowdisplaybreaks[4]

\newcommand{\be}{\begin{equation}}
\newcommand{\ee}{\end{equation}}
\newcommand{\beqa}{\begin{eqnarray}}
\newcommand{\eeqa}{\end{eqnarray}}

\renewcommand\l{\lambda}
\newcommand\m{\mu}
\newcommand\D{\Delta}
\newcommand\n{\nu}
\renewcommand\r{\rho}
\newcommand\s{\sigma}
\newcommand\vf{\varphi}
\renewcommand\a{\alpha}
\renewcommand{\b}{\beta}
\newcommand{\gc}{\mathpzc g}
\newcommand{\B}{B}

\def\d{\partial}

\newcommand{\bseq}{\begin{subequations}}
\newcommand{\eseq}{\end{subequations}}

\newcommand{\di}{\mathrm d}

\newcommand{\g}{\gamma}
\newcommand{\G}{\varGamma}
\newcommand{\z}{\zeta}
\newcommand{\bs}{\mbox{\boldmath $s$}}
\newcommand{\ve}{\varepsilon}
\newcommand{\bQ}{\mbox{\boldmath$Q$}}
\newcommand{\vS}{\varSigma}
\newcommand{\vO}{\varOmega}
\newcommand{\bG}{\mbox{\boldmath $\varGamma$}}
\newcommand{\vLam}{\varLambda}
\newcommand{\bUp}{\mbox{\boldmath $\varUpsilon$}}
\newcommand{\mg}{\mathfrak{g}}
\newcommand{\mN}{\mathfrak{N}}
\newcommand{\bq}{\mbox{\boldmath $q$}}

\begin{document}

\begin{titlepage}
\clearpage

\title{\vspace{-2cm} 
\begin{flushright}
{\normalsize
CERN-TH-2017-099}, 
%\vspace{-0.5cm}
{\normalsize INR-TH-2017-010}, 
{\normalsize FR-PHENO-2017-011}
\end{flushright}
\vspace{0.5cm} 
{\bf Renormalization of gauge theories in the background-field
  approach }}

\author{Andrei O.~Barvinsky$^{1,2}$, Diego Blas$^{3}$,
Mario Herrero-Valea$^{4}$,\\
Sergey M.~Sibiryakov$^{3,4,5}$, Christian F.~Steinwachs$^6$\\[2mm]
{\small\it $^1$ Theory Department, Lebedev Physics Institute, }%\\[-1mm]
{\small \it  Leninsky Prospect 53, Moscow 119991, Russia}\\
{\small \it $^2$ Department of Physics, Tomsk State University,
  Lenin Ave. 36, Tomsk 634050, Russia}\\
{\small\it $^3$ Theoretical Physics Department, CERN,
CH-1211 Geneva 23, Switzerland}\\
{\small\it $^4$ LPPC, Institute of Physics,
 EPFL, CH-1015 Lausanne, Switzerland}\\
{\small\it $^5$ Institute for Nuclear Research of the
Russian Academy of Sciences,}\\[-1mm]
{\small\it 60th October Anniversary Prospect, 7a, 117312
Moscow, Russia}\\
{\small\it $^6$ Physikalisches Institut,
  Albert-Ludwigs-Universit\"at
Freiburg,}\\[-1mm]
{\small\it Hermann-Herder-Strasse 3, 79104 Freiburg, Germany}
}

\date{}
\maketitle

\begin{abstract}
Using the background-field method we demonstrate the 
Becchi--Rouet--Stora--Tyu\-tin (BRST) structure  
of counterterms in a broad class of gauge theories. 
Put simply, we show that gauge invariance is preserved by
renormalization in local gauge field theories whenever they admit a sensible
background-field formulation and anomaly-free path integral measure.
This class encompasses
Yang--Mills theories (with possibly Abelian subgroups) 
and relativistic gravity, including  both renormalizable and
non-renormalizable (effective) theories. 
Our results also hold for non-relativistic models such as Yang--Mills theories with anisotropic scaling
or Ho\v rava gravity.  
They strengthen and generalize the existing results in the literature concerning the renormalization of
gauge systems.  
Locality of the BRST construction is emphasized
throughout the derivation. 
We illustrate our general approach with several explicit examples. 
\end{abstract}

\thispagestyle{empty}
\end{titlepage}

\newpage 
\tableofcontents

%%%%%%%%%%%%%%%%%%%%%%%%%%%%
\section{Introduction}
%%%%%%%%%%%%%%%%%%%%%%%%%%%%

A central question  in the perturbative quantization of gauge
field theories is to what extent the gauge symmetry is preserved by
renormalization. Intuition tells that in the absence of anomalies,
i.e. when the measure in the path integral  is gauge
invariant, the counterterms required to cancel the ultraviolet
divergences should be gauge invariant as well. A rigorous proof of
this assertion, however, is highly non-trivial due to 
 the breaking of the gauge
symmetry required to quantize gauge theories (gauge-fixing procedure). The original
gauge invariance still survives in the  
Becchi--Rouet--Stora--Tyutin 
(BRST) \cite{Becchi:1974md,Tyutin1} structure of the gauge-fixed
action $\vS$ which remains invariant under infinitesimal variations
generated by the nilpotent BRST operator. At tree level, 
$\vS$  is a sum of a BRST exact part responsible for
the gauge fixing and the
classical gauge invariant action which depends only on the physical
fields (gauge fields and matter) and is independent of the
Faddeev--Popov ghosts.  
The
physical content of gauge invariance will be retained if this BRST
structure persists under renormalization. In particular, it will
guarantee that the partition
function obeys Slavnov-Taylor identities at all orders of the
perturbative expansion. 

In addition, to preserve the key properties of quantum field theory,
the BRST structure must be compatible with locality. Namely, starting
from a gauge theory with a local\footnote{That is represented
  as a sum of terms depending on fields and their derivatives at a point.
This sum can, in principle, be
  infinite provided terms with higher number of derivatives are
  treated perturbatively, as it happens in effective field
  theories.} Lagrangian, both the BRST-exact and
the gauge-invariant parts of the renormalized action must be given by
integrals of local Lagrange densities.

In the textbook examples of renormalizable relativistic theories, such
as quantum electrodynamics or Yang--Mills (YM) theory, the previous properties
can be proven by ``brute force'': one first writes down all possible
counterterms allowed by power counting and then solves the equations
for their coefficients following from the Slavnov--Taylor
identities. The last step required to bring the renormalized action
into the BRST form is a field redefinition. Positive canonical
dimensions of the fields and the absence of any coupling constants with
negative dimensionality imply in these simple cases that the field
redefinition must have the form of a multiplicative wavefunction
renormalization, whose coefficient is easy to find, see
e.g.~\cite{Weinberg}. 

In general the situation is much more involved. This is the case, for example, 
in non-renormalizable theories (understood as effective field theories, EFTs) where one encounters coupling constants of negative 
dimension. 
In these cases an explicit
solution of the Slavnov--Taylor identities appears infeasible. Even if
such solution were available, the field redefinition bringing it to
the BRST form could be nonlinear and arbitrarily complicated, rendering
a brute-force search for it hopeless. The same is true for
renormalizable higher-derivative gravity~\cite{Stelle:1976gc}
where the canonical dimension of the metric is zero, implying
that its renormalization can be, and actually is, nonlinear. To study the
consistency of the BRST structure with renormalization in this type of
theories one needs more powerful methods.

It is well-known that the classification of possible counterterms
arising in general gauge theories requires computing the cohomology of
an extended BRST
operator \cite{ZinnJustin,Voronov:1982cp,Anselmi:1994ry,Gomis:1995jp}
(see also \cite{Weinberg}). To be compatible with the BRST structure,
the latter must consist of local\footnote{The requirement of locality
  is crucial. Refs.~\cite{Voronov:1982cp,Anselmi:1994ry} studying the
  BRST cohomology in general gauge theories do not guarantee its
  locality and have to postulate it as an additional assumption.}  
gauge-invariant functionals of 
physical fields only.
This was indeed
demonstrated in \cite{Barnich:1994ve,Barnich:1994mt} for the EFT consisting of
general relativity coupled to YM with semisimple gauge group extended
by arbitrary gauge invariant higher-order operators. These references
use the advanced mathematical apparatus of local cohomology theory. Notably,
for gauge groups with Abelian factors
they still leave room for non-trivial cohomologies different from
gauge invariant functionals which, if generated by divergences, would
imply deformations of the original gauge symmetry. Additional
arguments must be invoked to forbid the appearance of
such counterterms in the studied cases  \cite{Gomis:1995jp}.  

The purpose of our work is to address the BRST structure of
renormalized actions in general gauge field theories admitting
background field gauges. Our motivation is twofold. First, we
will provide a new, and we believe simpler, derivation of 
the results concerning the renormalization of 
Einstein--YM theories and strengthen them for the case of theories with Abelian
subgroups. Second, our analysis covers a broader class of 
gauge theories not considered in the classic papers
\cite{Barnich:1994ve,Barnich:1994mt}. This includes, in particular,
the higher-derivative 
gravity and gauge/gravity theories without relativistic
invariance. 

Non-relativistic gauge theories play a prominent role 
in 
condensed matter physics
\cite{Vafek:2002jf,Ardonne:2003wa,Roy:2015zna} (see also references therein), 
investigations of 
non-relativistic Weyl invariance and holography 
\cite{Kachru:2008yh,Griffin:2011xs},
and may be relevant for particle model building 
\cite{Anselmi:2008bt,Iengo:2010xg} (see \cite{Barvinsky:2017mal} for a summary of extra motivations and results in non-relativistic gauge theories). 
Furthermore, abandoning relativistic invariance (while keeping the gauge group of
time-dependent spatial diffeomorphisms) allows one to construct 
power-counting renormalizable
models of gravity in arbitrary spacetime dimensions including the
phenomenologically interesting case of 
$(3+1)$ dimensions \cite{Horava:2008ih}. The renormalizability beyond
power counting was established in \cite{Barvinsky:2015kil}  for a
large subset of these gravity models, the so-called
projectable Ho\v rava gravities. 
It was assumed in \cite{Barvinsky:2015kil} that renormalization
preserves gauge invariance, which was explicitly demonstrated only at
one loop. One of the goals of the present paper is to demonstrate the
validity of this assumption to all loop orders and thereby complete
the proof of renormalizability of projectable Ho\v rava gravity.
 
Our approach is based on the background field method
\cite{DeWitt:1967ub,Abbott:1981ke} (see also 
\cite{DeWitt_book,Veltman}), a powerful tool for calculating  the
quantum effective action in gauge
theories and gravity. The main virtue of this method is that it  preserves
the gauge invariance of the calculations even after gauge fixing. 
This is 
achieved by the introduction of additional external
sources --- background fields --- in such a way that the partition
function remains invariant under {\em simultaneous} gauge
transformations of the variables in the path integral (``quantum
fields'') and the background fields. We denote this transformation ``background-gauge transformation''. At the same time the {\em
  quantum} gauge
transformations acting only on the quantum fields are broken by
gauge fixing and the path integral is well
defined (at least perturbatively). The construction of  {\em background-covariant} gauge fixing
conditions is straightforward in 
theories containing fields in 
linear representations of gauge groups with linear
generators. 
These conditions imply that the background-gauge symmetry is preserved by
renormalization which serves as a strong selection criterion for 
possible counterterms. This method greatly simplifies the renormalization of
coupling constants in the one-loop approximation after the  background fields are identified with 
the mean value of  the quantum fields \cite{Honerkamp:1972fd,tHooft:1973bhk,tHooft:1974toh}. In this case  
the counterterms take a manifestly gauge
invariant form.  

Beyond one-loop the situation becomes more
complicated. The subtraction of subdivergences necessary to eliminate the
nonlocal  infinities requires counterterms where the quantum fields are
distinct from the background fields. Background-gauge invariance is
not sufficient to completely fix the structure of such counterterms 
and the BRST structure associated to the quantum gauge
transformations must be exploited, as is done in the cases of gauges
without 
background fields 
\cite{KlubergStern:1974xv,Tyutin:1978ec,Grassi:1995wr,
Ferrari:2000yp,Binosi:2011ar}.  
In practical calculations these counterterms can sometimes be avoided by 
subtle
methods that have been developed for YM and relativistic gravity.
However, these techniques  
generically feature nonlocal divergences at intermediate steps of
the calculations, that cancel only in the final quantities evaluated 
on-shell \cite{Kallosh:1974yh,Arefeva:1974jv,Abbott:1980hw,Ichinose:1981uw,
Barvinsky-Vilkovisky}. The presence of nonlocal divergences makes these methods
inappropriate for a general analysis of renormalizability. 
More recently it has been advocated \cite{Anselmi:2013kba} that the
use of a background gauge combined with the standard subtraction scheme
provides a valuable tool for such analysis (see also \cite{Binosi:2012pd}). 
This reference uses the
Batalin--Vilkovisky formalism \cite{Batalin:1981jr,BV3} to prove the
existence of a canonical transformation bringing the renormalized
action to the BRST form. However, this requires introducing
background field counterparts for all quantum fields of the theory
including Faddeev--Popov ghosts and, moreover, the addition of
Batalin--Vilkovisky antifields for all background fields. Such
proliferation of objects makes the construction rather baroque and
obscures the subtleties of the derivation.
 
In this paper we adopt a different strategy and proceed along the
lines of traditional cohomology analysis. Our key finding is that the
background-gauge invariance greatly facilitates the computation of the
local BRST cohomology. The latter reduces to cohomologies of a few
simpler nilpotent operators that are readily computed using elementary
algebraic techniques. The resulting constraints on the form of the
renormalized action imply that, upon an appropriate field
redefinition, it acquires the desired BRST form (a local gauge-invariant
functional plus a BRST-exact piece). 
The argument does not involve any power-counting
considerations. When available, such considerations lead to further
refinements which we discuss.
We keep track of locality at all steps of the
derivation. 

Our proof applies to theories characterized by the following
properties: the gauge algebra is
irreducible and closes off-shell;  the gauge generators 
depend on the fields at most linearly; the structure functions are 
field independent. These conditions ensure that the theory admits a
convenient background-covariant gauge fixing.
Besides, we assume the absence of anomalies and 
locality of the leading ultraviolet divergences (ones that
  remain after subtraction of subdivergences).
The latter requirement should not be confused with locality of the
BRST decomposition, which is not postulated a priori, but is derived
from the previous assumptions.
 
The above class is quite broad. 
It encompasses renormalizable and non-renormalizable (effective)
theories with Abelian and non-Abelian gauge groups, 
general relativity and higher-derivative
gravity. Besides the standard relativistic versions of these theories,
it also includes  
their non-relativistic generalizations 
\cite{Anselmi:2008bq,Horava:2008jf,Horava:2008ih}. 
As a corollary of our general result we establish for the first time
the compatibility of the BRST structure with renormalization in
projectable Ho\v rava gravity \cite{Horava:2008ih} which completes the
proof of its renormalizability. 
A notable
example that is not covered by our study is supergravity where the gauge algebra
closes only on-shell\footnote{For $N=1$ supergravity in four spacetime
  dimensions, the off-shell closure of the algebra can be
  achieved by introduction of auxiliary fields, but then the
  generators become nonlinear in the fields \cite{WessBagger}.}.  

While various ingredients of our analysis have already appeared in the
literature, to the best of our knowledge, they have never been put
together. To make the presentation self-contained we review these
ingredients in the relevant sections. Several concrete examples aim to
illustrate the physical content of the general result. 
For simplicity we focus throughout the paper on theories with bosonic gauge
parameters.    

The paper is organized as follows. In Sec.~\ref{sec:2} we describe our
assumptions, introduce the background
gauge fixing and formulate our main result (Sec.~\ref{sec:2.4}). 
In
Sec.~\ref{sec:examples} we illustrate its implications on several
examples.  
We discuss explicitly the standard
renormalizable YM in $(3+1)$ dimensions, relativistic higher-derivative gravity in  $(3+1)$ dimensions, projectable
Ho\v rava gravity  in general dimensions and general relativity  in
$(3+1)$ dimensions (understood as an effective theory). In
Sec.~\ref{sec:4} 
we turn to the proof of our general result and derive
the equations satisfied by the effective action as a
consequence of the
background and quantum gauge invariances. 
These equations are used to analyze the structure of
the divergent counterterms in Sec.~\ref{sec:5},
which is the central part of the paper.
Here we formulate the cohomology problem and use the background-gauge
invariance to split it into several subproblems. Solving them we fix
the structure of the renormalized action and demonstrate existence of
a field redefinition that casts it into the BRST form advocated in
Sec.~\ref{sec:2.4}. This completes the formal proof.  
Section~\ref{sec:woex} is devoted to one more example ---
the $O(N)$ vector model in $(1+1)$ spacetime dimensions written as an
Abelian gauge theory. This example is interesting as it features
nonlinear wavefunction renormalization, being at the same time simple
enough to admit an explicit treatment.
We verify at one loop that the 
counterterms
in this theory have the structure determined by the general argument. 
We conclude in Sec.~\ref{sec:conclusions}. 
Appendix~\ref{app:STW} contains the derivation of
the Slavnov--Taylor and Ward identities for the partition function.
In Appendix~\ref{app:coh} we prove a lemma about 
the cohomology of an operator appearing in our analysis.
Some formulae used in 
the computation of the effective action of the $O(N)$
model are summarized in 
Appendix~\ref{app:sigtech}.

%%%%%%%%%%%%%%%%%%%%%%%%%%%%
\section{Assumptions and proposition}
\label{sec:2}
%%%%%%%%%%%%%%%%%%%%%%%%%%%%

%%%%%%%%%%%%%%%%%%%%%%%%%%%%
\subsection{Gauge algebra}
\label{sec:2.1}
%%%%%%%%%%%%%%%%%%%%%%%%%%%%

We consider a
theory with local gauge and matter fields $\varphi^a$, 
where $a$ is a collective
notation for all indices and the coordinates. 
The theory is described by the action $S[\,\varphi\,]$ which is an
integral of a local Lagrangian density ${\cal L}(\vf)$. The latter is
expanded as a sum of terms depending on the fields $\vf^a$ and their
finite-order derivatives at a given point\footnote{Throughout the text
the dependence of local functions on the fields and their finite-order
derivatives will be denoted by round brackets, while square brackets
will denote the functional
dependence of integral quantities with local or nonlocal integrands.}.  
The action $S[\vf]$ is
invariant under gauge
transformations with local bosonic parameters $\ve^\a$. The
transformations are assumed to
have at most linear dependence on the fields, 
\begin{align}
\label{gtrans}
\delta_\ve \varphi^a = R^a_{~\a}(\varphi)\,\ve^\a, \quad
R^a_{~\a}(\varphi)=P^a_{~\a}+R^a_{~b\a}\varphi^b\;,\quad R^a_{~\a}(\varphi)\,\frac{\delta S[\,\varphi\,]}{\delta\varphi^a}=0\;.
\end{align}
We further assume that the gauge algebra closes off-shell,
\be
\label{closure}
\big[\delta_\ve,\delta_\eta\big]\varphi^a=\delta_\varsigma \varphi^a\;,
\ee
where
\be
\label{structC}
\varsigma^\a=C^\a_{~\b\g}\ve^\b\eta^\g \, ,
\ee
and $C^\a_{~\b\g}$ are field-independent structure functions. 
The closure condition
implies the relations,
\bseq
\label{compat*}
\begin{align}
\label{compat1}
&R^a_{~b\a}P^b_{~\b}-R^a_{~b\b}P^b_{~\a}=P^a_{~\g}C^\g_{~\a\b}\;,\\
\label{compat2}
&R^a_{~b\a}R^b_{~c\b}-R^a_{~b\b}R^b_{~c\a}=R^a_{~c\g}C^\g_{~\a\b}\;.
\end{align}
\eseq
Besides, $C^\a_{~\b\g}$ obey the Jacobi identities,
\be
\label{Jacobi}
C^\a_{~\b[\g}C^\b_{~\l\m]}=0\;,
\ee
where the square brackets mean anisymmetrization over the respective
indices.

Next, we require that the set of gauge generators $R^a_{~\a}(\vf)$ is
{\em locally complete} and {\em irreducible}. These properties are defined as
follows:
\begin{enumerate}
 \renewcommand{\theenumi}{(i)}
  \renewcommand{\labelenumi}{{\theenumi}}
\item \label{(i)} \emph{Local completeness}~\cite{BV3,Henneaux:1990rx}: 
Any local operator $X^a_{~\a}(\vf)$
  satisfying the equation
\be
\label{Sannih}
\frac{\delta S}{\delta\vf^a} X^a_{~\a}=0,
\ee
is represented as a linear combination of the gauge generators and
equations of motion,
\be
\label{Xrepr}
X^a_{~\a}=R^a_{~\b}\, Y^\b_\a+\frac{\delta S}{\delta\vf^b}\,I^{[ba]}_{\a}\,,
\ee
where $Y^\b_\a$ and $I^{[ba]}_\a$ are {\em local} and $I^{[ba]}_\a$ is antisymmetric
in its indices. The locality condition means that $Y^\b_\a$ and
$I^{[ba]}_\a$ are non-zero only if the coordinates corresponding to
$\b$ and $\a$ or $a$, $b$ and $\a$ coincide.

 \renewcommand{\theenumi}{(ii)}
  \renewcommand{\labelenumi}{{\theenumi}}
\item\label{(ii)}  \emph{Irreducibility~}\cite{Batalin:1981jr}: 
Let $\vf^a_0$ be a solution of the
  equations of motion, so that  
\be
\label{eoms}
\frac{\delta S}{\delta\vf^a}(\vf_0)=0\;.
\ee
If a gauge parameter $\ve^\a$ satisfies the relations
\be
\label{Rannih}
R^a_{~\a}(\vf_0)\,\ve^\a=0\;,
\ee
then $\ve^\a=0$. In other words, gauge transformations
act non-trivially on  on-shell configurations.
\end{enumerate}

The class of theories described above is quite broad. 
It includes, in particular,  
relativistic Abelian and non-Abelian gauge theories together with their
extensions by higher-derivative operators, general
relativity and relativistic higher-derivative
gravity, e.g.~\cite{Stelle:1976gc}. Besides, it contains 
non-relativistic generalizations of these theories.
Some examples are discussed in 
Sec.~\ref{sec:examples} and in
Sec.~\ref{sec:woex}. As we mentioned, a notable
exception from this class
is supergravity, both due to the fermionic nature of the gauge
parameter and openness of the gauge algebra.

For the sake of clarity, we focus in what follows on theories where
all fields $\vf^a$ are bosonic. The inclusion of fermionic {\em matter} fields
is straightforward, but would complicate the formulae by additional
$(-1)$ factors.

%%%%%%%%%%%%%%%%%%%%%%%%%%%%
\subsection{Background gauge}
\label{sec:gauge}
%%%%%%%%%%%%%%%%%%%%%%%%%%%%

To quantize the theory we need to fix the gauge.
We introduce the {\em background fields}\footnote{For bosonic gauge
  algebras that we consider in this paper, 
it is sufficient to
  introduce background counterparts to bosonic fields only, even if
  the theory contains fermionic matter.} $\phi^a$ and
choose the gauge fixing function $\chi^\a(\varphi,\phi)$ in such a way that
it transforms covariantly under simultaneous local gauge transformation of
$\varphi^a$ and $\phi^a$ with the same parameter $\varepsilon$
but their own generators $R^a_{~\a}(\varphi)$ and $R^a_{~\a}(\phi)$
respectively, 
\be
\label{btrans}
\delta_\ve\varphi^a=R^a_{~\a}(\varphi)\,\ve^\a~,~~~
\delta_\ve\phi^a=R^a_{~\a}(\phi)\,\ve^\a\;.
\ee
Covariance of $\chi^\a$ under the transformations
(\ref{btrans}) implies,
\be
\label{chitrans}
\delta_\ve\chi^\a
\equiv\frac{\delta\chi^\a}{\delta\vf^a}\delta_\ve\varphi^a
+\frac{\delta\chi^\a}{\delta\phi^a}\delta_\ve\phi^a
=-C^\a_{~\b\g}\chi^\b\ve^\g\;.
\ee
We will refer to (\ref{btrans}) as ``background-gauge
transformations'' and to $\chi^\a(\varphi,\phi)$ as 
``back\-gro\-und-covariant gauge conditions''.
We further choose $\chi^\a$ to be linear in the difference
$(\varphi^a-\phi^a)$, 
\be
\label{chilin}
\chi^\a (\varphi,\phi)=\chi^\a_a(\phi)\,(\varphi^a-\phi^a)\;.
\ee
The gauge-fixing function is assumed to be local in
space-time, i.e. it depends only on the values of the fields and their
derivatives of finite order at a point. 

The gauge fixing is implemented by the BRST 
procedure \cite{Becchi:1974md,Tyutin1} (see also \cite{Weinberg}). 
Labelling anticommuting ghosts $\omega^\a$, antighosts
$\bar\omega_\a$ and the Lagrange multiplier $b_\a$ with the condensed
gauge index $\alpha$, we define the standard action of the BRST
operator $\bs$
\bseq
\label{BRST*}
\begin{align}
\label{BRST1}
&\bs \varphi^a=R^a_{~\a}(\varphi)\,\omega^\a\;,\\
\label{BRST2}
&\bs\, \omega^\a=\frac{1}{2}C^\a_{~\b\g}\,\omega^\b\omega^\g\;,\\
\label{BRST3}
&\bs\, \bar\omega_\a=b_\a\;,\\
\label{BRST4}
&\bs b_\a=0\;.
\end{align}
\eseq
The closure conditions (\ref{compat*}), (\ref{Jacobi})
imply that $\bs$ is nilpotent. 
The background fields $\phi^a$ are invariant under the action of
$\bs$. 
Next, we introduce two sets of anticommuting
auxiliary fields $\g_a$, $\varOmega^a$ and a commuting field $\z_\a$. 
They are also invariant
under the BRST transformations generated by $\bs$. We define
the {\em gauge fermion} as
\be
\label{Gauge_fermion}
\varPsi_0[\,\varphi,\omega,\bar\omega,b,\phi,\g,\z\,]=
\bar\omega_\a\bigg(\chi^\a_a(\phi)(\varphi^a-\phi^a)-
\frac{1}{2}O^{\a\b}(\phi)\,b_\b\bigg)
-\g_a(\varphi^a-\phi^a)+\z_\a\omega^\a\;.
\ee
Here $O^{\a\b}(\phi)$ is an invertible local operator that can, in general,
depend on the background fields\footnote{This dependence  is, in
  fact, inevitable in gravity (see Sec.~\ref{sec:examples}).} 
and transforms covariantly under the
background-gauge transformations. 
Finally, we construct the gauge-fixed action,
\be
\label{Sigshort}
\varSigma_0[\vf,\omega,\bar\omega,b,\phi,\g,\z,\varOmega]=S[\vf]+\bQ\,\varPsi_0\;,
\ee
with
\be
\label{Qlong}
\bQ=\bs+\varOmega^a\frac{\delta}{\delta\phi^a}\;.
\ee
Following \cite{KlubergStern:1974xv,Tyutin:1978ec,Grassi:1995wr,
Ferrari:2000yp,Binosi:2011ar} we have extended the usual BRST operator
in such a way that it 
controls not only the 
field BRST transformations but also the variation of the gauge-fixing
term under the changes of $\phi$. Clearly, $\bQ$ is nilpotent due to
the anticommuting nature of $\vO^a$.
Explicitly, the action (\ref{Sigshort}) reads,
\be
\label{Sig}
\begin{split}
\varSigma_0[\varphi,\omega,\bar\omega,b,\phi,\g,\z,\varOmega]
=S[\,\varphi\,]+b_\a\chi^\a_a(\phi)\,(\varphi^a-\phi^a)-
\frac{1}{2}O^{\a\b}(\phi)\,b_a b_\b
-\bar\omega_\a\,\chi^\a_a(\phi)\,R^a_{~\b}(\varphi)\,\omega^\b\\
+\g_aR^a_{~\a}(\varphi)\,\omega^a
+\frac12\,\z_\a C^\a_{~\b\g}\,\omega^\b\omega^\g\;
+\varOmega^c\,\bar\omega_\a
\bigg[\frac{\delta\chi^\a_b}{\delta \phi^c}(\varphi-\phi)^b-\chi^\a_c-
\frac{1}{2}\frac{\delta O^{\a\b}}{\delta \phi^c}b_\b\bigg]
+\varOmega^c\gamma_c\;.
\end{split}
\ee
One recognizes the gauge fixing part (second and third terms in the
first line)\footnote{Gaussian integration over the Lagrange multiplier
  $b_\a$ gives a familiar gauge breaking term $\frac12\,\chi^\a
  O^{-1}_{\a\b}\chi^\b$ with the weighting factor inverse to
  $O^{\a\b}$.} and the Faddeev--Popov action for the ghost-antighost pair
(last term in the first line). The second line collects the dependence on the
auxiliary fields $\g_a$, $\z_\a$ and $\varOmega^a$. Notice that $\g_a$
and $\z_\a$ couple as sources to the BRST variations of $\vf^a$ and
$\omega^\a$ respectively. 

In view of the nilpotency of $\bQ$ the gauge-fixed action is BRST-invariant, 
\be
\label{BRSTS0}
\bQ \varSigma_0=0\;.
\ee
This equation will be used below to derive the Slavnov--Taylor
identities constraining the ultraviolet 
divergences. 
Besides,
for background-covariant gauges of the above type, $\varPsi_0$ and
$\varSigma_0$
have an additional
symmetry: 
they are invariant under background-gauge transformations~(\ref{btrans}),
\be
\label{binvar}
\delta_\ve\varPsi_0=0,\quad \delta_\ve\varSigma_0=0\,,
\ee
if simultaneously with $\vf^a$ and $\phi^a$ we transform
all fields in the appropriate
linear representations:
\be
\label{btrans1}
\delta_\ve\g_a=-\g_bR^b_{~a\a}\ve^\a~,~~~
\delta_\ve\omega^\a=-C^\a_{~\b\g}\,\omega^\b\ve^\g~,~~~
\delta_\ve\z_\a=\z_\b C^\b_{~\a\g}\ve^\g,~~~
\delta_\ve\varOmega^\a=R^a_{~b\a}\varOmega^b\ve^\a\;,
\ee
and similarly for $\bar\omega_\a$ and $b_\a$.  
Note that for theories
with diffeomorphism invariance $\omega^\a$ transforms as a
contravariant vector, whereas $\bar\omega_\a$, $b_\a$, $\g_a$, $\z_\a$ are
vector/tensor densities. 
Finally, the action (\ref{Sig}) possesses a global $U(1)$ symmetry
corresponding to the ghost number with the following assignment of charges:
\be
\label{Qgh}
\begin{split}
&{\rm gh}(\varphi)={\rm gh}(\phi)={\rm gh}(b)=0~,~~~
{\rm gh}(\omega)={\rm gh}(\varOmega)=+1\;,\\
&{\rm gh}(\bar\omega)={\rm gh}(\g)=-1~,~~~{\rm gh}(\z)=-2\;.
\end{split}
\ee

Using (\ref{Sig}) as the tree-level action and
introducing sources coupled to the ``quantum'' fields
$(\vf,\omega,\bar\omega,b)$ we  write the ``bare''
generating functional for connected graphs, 
\be
\label{W0}
W_0[J,\bar\xi,\xi,y,\phi,\g,\z,\varOmega]
=-\hbar\log\int d\varPhi
\exp\bigg[-\frac{1}{\hbar}\big(\varSigma_0+J_a(\varphi^a-\phi^a)
+\bar\xi_\a\omega^\a+\xi^\a\bar\omega_\a
+ y^\a b_\a\big)\bigg]\,.
\ee
Here we have collectively denoted all quantum fields 
by 
$\varPhi$ in the integration measure and explicitly
included the Planck constant $\hbar$ as a counting parameter for the
order of the  loop 
expansion\footnote{Throughout the paper we work with Euclidean field
  theory and use the corresponding definition of the generating functional.
As the
  operator $O^{\a\b}$ in (\ref{Sig}) is usually chosen
  positive-definite, the convergence of the path integral requires
  that the integration in $b_\a$ runs along the imaginary axis. This
  subtlety does not affect our analysis.}. 

%%%%%%%%%%%%%%%%%%%%%%%%%%%%
\subsection{Absence of anomalies and locality of divergences}
\label{sec:2.3}
%%%%%%%%%%%%%%%%%%%%%%%%%%%%

We impose two more conditions on the theory. First, we  postulate  the absence
of gauge anomalies, i.e. the existence of a regularization prescription
that preserves the gauge invariance of the functional integration
measure. This is achieved by dimensional regularization in many cases. 

Second, we require that a variant of the standard
subtraction
scheme (e.g. minimal subtraction) \cite{Collins} 
eliminates all nonlocal divergences. 
Let us expand on this point. In the standard
scheme the counterterms are constructed inductively in the number of
loops $L$ or, equivalently, in the powers of $\hbar$. Let us assume that at
order ${\cal O}(\hbar^{L-1})$ we have already constructed the
renormalized action 
\be
\label{SigmaL-1}
\varSigma_{L-1}=\varSigma_0+\sum_{l=1}^{L-1}\hbar^l\varSigma^C_l\;,
\ee
where $\varSigma_0$ is the tree-level action (\ref{Sig}) and 
$\varSigma^C_l$ are divergent local counterterms. This action is such that 
the 
generating functional $W_{L-1}$ defined by the formula analogous
to (\ref{W0}) with the replacement
$\varSigma_0\mapsto\varSigma_{L-1}$
produces Green's functions that are finite at $(L-1)$ loops.

Next,  we introduce the mean fields\footnote{These should not be
  confused with
  the background fields $\phi^a$.} as functional derivatives of the
generating functional with respect to the sources\footnote{
We fix the sign of the derivatives with respect to the anticommuting
variables by placing the differential on the left, $df=d\theta
f'(\theta)$.},
\be
\label{avfields}
\langle\varphi^a\rangle-\phi^a=\frac{\delta W}{\delta J_a}~,~~~
\langle\omega^\a\rangle=\frac{\delta W}{\delta \bar\xi_\a}~,~~~
\langle\bar\omega_\a\rangle=\frac{\delta W}{\delta \xi^\a}~,~~~
\langle b_\a\rangle=\frac{\delta W}{\delta y^\a}\;,
\ee
and define the effective action $\varGamma$ as
the Legendre transform of $W$,
\be
\label{Gamma}
\varGamma\big[\langle\varphi\rangle,\langle\omega\rangle,
\langle\bar\omega\rangle,\langle b\rangle,\phi,
\g,\z,\varOmega\big]=
W-J_a(\langle\vf^a\rangle-\phi^a)
-\bar\xi_\a\langle\omega^\a\rangle
-\xi^\a\langle\bar\omega_\a\rangle
-y^\a\langle b_\a\rangle\;.
\ee
Clearly, it satisfies,
\be
\label{sources}
\frac{\delta\G}{\delta\langle\varphi^a\rangle}=-J_a~,~~~
\frac{\delta\G}{\delta\langle\omega^\a\rangle}=\bar\xi_\a~,~~~
\frac{\delta\G}{\delta\langle\bar\omega_\a\rangle}=\xi^\a~,~~~
\frac{\delta\G}{\delta\langle b_\a\rangle}=-y^\a\;,
\ee
The $(L-1)$-th order effective action has the
form
\be
\label{GL-1}
\G_{L-1}=\varSigma_0+\sum_{l=1}^\infty \hbar^l\G_{L-1}^{(l)}\;,
\ee
where $\G_{L-1}^{(l)}$ is the contribution of diagrams with $l$ loops.
By the assumption of the induction step, 
all terms $\G^{(l)}_{L-1}$ with $l\leq L-1$ 
are finite and the divergence of the $L$-th term,
\be
\label{localdiv}
\G^{(L)}_{L-1,\infty}\equiv\G_{L,\infty}
[\langle\varphi\rangle,\langle\omega\rangle,\langle\bar\omega\rangle,
\langle b\rangle,\phi,\g,\z,\varOmega]
\ee
is local. Then the counterterm $\varSigma_L^C$ is identified with
$-\G_{L,\infty}$ where the mean fields are replaced by the quantum
fields,
\be
\label{SigmaL}
\varSigma_L[\vf,\omega,\bar\omega,b,\phi,\gamma,\z,\Omega]=
\varSigma_{L-1}-\hbar^L\G_{L,\infty}
[\varphi,\omega,\bar\omega,b,\phi,\g,\z,\varOmega]\;.
\ee
According to the standard theorems
\cite{Collins} 
(see \cite{Anselmi:2007ri}
for the generalization to theories without Lorentz invariance), this
subtraction removes the $L$-loop divergences, as well as all
subdivergences in $(L+1)$-loop diagrams. 

In relativistic gauge theories
with Lorentz-covariant gauge fixing,
this guarantees that the remaining divergence of order ${\cal
  O}(\hbar^{L+1})$ in the effective action $\G_{L}$ is
local and the subtraction can be repeated at the $(L+1)$-th loop
order.  The situation is less straightforward in the absence of Lorentz
invariance \cite{Barvinsky:2015kil} and the locality of the remaining
divergences must be verified in every given theory. It was shown to
hold for non-relativistic YM theories with anisotropic (Lifshitz) scaling and
projectable Ho\v rava gravity \cite{Barvinsky:2015kil}. In the
present paper we postulate it as a property of the class of theories under study.

To avoid cluttered notations, we will omit the averaging symbols on the
arguments of the effective action $\G$ in what follows.

%%%%%%%%%%%%%%%%%%%%%%%%%%%%
\subsection{Proposition: BRST structure of the renormalized action}
\label{sec:2.4}
%%%%%%%%%%%%%%%%%%%%%%%%%%%%

We will show that a slight modification of the subtraction
prescription by the inclusion of additional local terms of order 
${\cal O}(\hbar^{L+1})$ on the r.h.s. of (\ref{SigmaL}) leads to a
renormalized action $\vS_L$ that preserves the BRST structure.
More precisely, our result is formulated as follows.

Let us denote the fields coupled to the external sources
$J,\bar\xi$ by $\tilde\vf,\tilde\omega$ and consider local 
 field reparameterizations of the form,
\be
\label{reparam}
\tilde\vf^a=\tilde\vf_L^a(\vf,\omega,\phi,\hat\gamma,\z,\vO)\,~~~~
\tilde\omega^\a=\tilde\omega^\a_L(\vf,\omega,\phi,\hat\gamma,\z,\vO)\;,
\ee
where we have introduced the combination
\be
\label{gahat}
\hat\g_a=\g_a-\bar\omega_\a\chi^\a_a(\phi)\;    
\ee
that will play an important role below. Upon the field redefinition the $L$-th order
generating functional reads\footnote{We disregard the functional Jacobian
  $|\delta\tilde\varPhi/\delta\varPhi|$ which gives an ultralocal
  contribution to the action. Such contributions vanish in dimensional
  regularization.}, 
\be
\label{Wgen}
W_L[J,\bar\xi,\xi,y,\phi,\g,\z,\varOmega]
=-\hbar\log\int d\varPhi
\exp\bigg[-\frac{1}{\hbar}\bigg(\varSigma_L+J_a(\tilde\vf^a_L-\phi^a)
+\bar\xi_\a\tilde\omega^\a_L+\xi^\a\bar\omega_\a
+y^\a b_\a\bigg)\bigg].
\ee
We will demonstrate the existence of a field redefinition (\ref{reparam})
such that $\vS_L$ takes the form,
\be
\label{BRSTSL}
\vS_L[\vf,\omega,\bar\omega,b,\phi,\gamma,\zeta,\vO]=S_L[\vf]+\bQ\,
\varPsi_L[\vf,\omega,\bar\omega,b,\phi,\gamma,\z,\vO]\;,
\ee
where $S_L[\vf]$ is a gauge invariant \emph{local} functional and
the BRST operator $\bQ$ has been defined in (\ref{Qlong}).
The gauge fermion
$\varPsi_L$ is a \emph{local} functional with ghost number $(-1)$ which is 
invariant under background-gauge transformations (\ref{btrans}),
(\ref{btrans1}) and has the form,
\be
\label{Psigen}
\varPsi_L=\hat\varPsi_L[\vf,\omega,\phi,\hat\gamma,\z,\vO]
-\frac12\, \bar\omega_\a O^{\a\b}(\phi)b_\b\;,
\ee 
where
\be
\label{Psihattree}
\hat\varPsi_L=-\hat\g_a(\vf^a-\phi^a)+\z_\a\omega^\a+{\cal O}(\hbar)\;.
\ee
Further, the reparameterization (\ref{reparam}) itself
is generated by the gauge fermion,
\be
\label{reparam1}
\tilde\vf^a_L=\phi^a
-\frac{\delta\varPsi_L}{\delta\g_a}\;,~~~~
\tilde\omega^\a_L=
\frac{\delta\varPsi_L}{\delta\z_\a}\;.
\ee
Together with (\ref{Psihattree}) this implies that at tree level 
$\tilde\vf,\tilde\omega$ coincide with $\vf,\omega$ and the gauge
fermion $\varPsi_L$ coincides with the expression
(\ref{Gauge_fermion}). Thus, we recover \eqref{W0} at tree level. 

Eqs.~(\ref{Wgen})---(\ref{reparam1}) represent a generalization of
the construction described in Sec.~\ref{sec:gauge} that is forced on us
by renormalization.
Notice that the sources $J_a$, $\bar\xi_\a$ now couple to composite
local operators that in general depend not only on the quantum fields,
but also on the external backgrounds $\phi^a$, $\g_a$, $\z_\a$,
$\vO^a$. Nevertheless, this is not problematic due to the property
(\ref{Psihattree}), (\ref{reparam1}) that ensures linearity of the coupling at 
leading order in $\hbar$.  

We will see in Sec.~\ref{sec:examples} and Sec.~\ref{sec:woex} that in many
interesting theories, that are typically renormalizable, power counting
considerations strongly restrict the dependence of the \emph{renormalized} gauge
fermion $\hat\varPsi_L$ on the auxiliary fields. Namely, in these cases
$\hat\varPsi_L$ is independent of $\vO^a$ and can depend on
$\hat\g_a$, $\z_\a$ only linearly,
\be
\label{Psirenorm}
\hat\varPsi_L=-\hat\g_a U^{~a}_L(\vf,\phi)+\z_\a\omega^\b
V^{~\a}_{L\b}(\vf,\phi)\;,
\ee
with
\bseq
\label{UVL}
\begin{align}
\label{UL}
&U_L^{~a}=\vf^a-\phi^a+\sum_{l=1}^L\hbar^l\mbox{\boldmath
  $u$}_l^{\;a}(\vf,\phi)\;,\\
\label{VL}
&V_{L\b}^{~\a}=\delta_\b^\a+\sum_{l=1}^L\hbar^l\mbox{\boldmath
  $v$}_{l\b}^{\;\a}(\vf,\phi)\;.
\end{align}
\eseq
Correspondingly, the field redefinition (\ref{reparam1}) bringing the
counterterms into the BRST-invariant form simplifies to
\be
\label{reparamrenorm}
\tilde\vf^a_L=\phi^a+U^{~a}_L(\vf,\phi)~,~~~~~
\tilde\omega^\a_L=V^{~\a}_{L\b}(\vf,\phi)\,\omega^\b\;.
\ee  
In this case it does not involve the auxiliary sources
$\g_a,\z_\a,\vO^a$.

%%%%%%%%%%%%%%%%%%%%%%%%%%%%
\section{BRST structure for selected gauge theories}
\label{sec:examples}
%%%%%%%%%%%%%%%%%%%%%%%%%%%%

In this section we illustrate the notions and results described above on several
gauge theories and discuss restrictions imposed on the structure
of divergences by power counting in renormalizable cases.
Together with a few well-known examples we consider the case of
projectable Ho\v rava gravity whose BRST structure is studied here for
the first time. 
 Readers interested
in the general proof can skip this section and 
proceed directly to Sec.~\ref{sec:4}.

%%%%%%%%%%%%%%%%%%%%%%%%%%%%
\subsection{Relativistic Yang--Mills in $(3+1)$ dimensions}
\label{sec:YM}
%%%%%%%%%%%%%%%%%%%%%%%%%%%%

As a first example, we consider the standard YM theory 
in $(3+1)$ spacetime dimensions. It has been already studied using
an approach similar to ours in 
\cite{KlubergStern:1974xv,Tyutin:1978ec,Grassi:1995wr,
Ferrari:2000yp,Binosi:2011ar}. Let us start by
expanding the condensed notations\footnote{Where no confusion is possible, we keep a condensed notation for space-time coordinates as $x$ and 
their delta functions as $\delta(x)$.},
\bseq
\label{YMfields}
\begin{gather}
\vf^a\mapsto A_\m^i(x)~,~~~~~
\varepsilon^\a\mapsto \varepsilon^i(x)\;,\\
R^a_{~b\a}\mapsto f^{ijk}\,\delta_\m^\n\,\delta(x-x_1)\,\delta(x-x_2)~,~~~~~
P^a_{~\a}\mapsto \delta^{ij}\,\d_\m\delta(x-x_1)\;,\\
C^\a_{~\b\g}\mapsto f^{ijk}\,\delta(x-x_1)\,\delta(x-x_2)\;,
\end{gather}
\eseq
where $A_\m^i(x)$ is the usual Yang--Mills field, $i$ is the color
index and $f^{ijk}$ are the totally antisymmetric coordinate independent
structure constants of the gauge group. We next introduce the
background field $\B_\m^i(x)$ and the gauge-fixing function,
\be
\label{YMgauge}
\chi^\a\mapsto \d^\m(A^i_\m-\B^i_\m)+f^{ijk}\B^{j\m}
(A^k_\m-\B^k_\m)\equiv D_{(\B)}^\m (A^i_\m-\B^i_\m)\;.
\ee 
Introducing the Faddeev--Popov ghosts $\omega^i(x)$, antighosts 
$\bar\omega^i(x)$,
the Lagrange multiplier $b^i(x)$
and the BRST sources 
\be
\label{YMaux}
\g_a\mapsto \g^{i\m}(x)~,~~~~\z_\a\mapsto\z^i(x)
~,~~~~\vO^a\mapsto \vO^i_\m(x)\;,
\ee
we obtain the gauge-fixed action,
\be
\label{YMSig}
\begin{split}
\vS_0=\int {\rm d}^4x\bigg[\frac{1}{4{\gc}^2}F_{\m\n}^iF^{i\m\n}
&+b^iD_{(\B)}^\m (A_\m^i-\B_\m^i)-\frac{\a}{2}b^ib^i
+D_{(\B)}^\m\bar\omega^i D_{(A)\m}\omega^i\\
&+\g^{i\m}D_{\m(A)}\omega^i+\frac{1}{2}\z^if^{ijk}\omega^j\omega^k
+\vO^i_\m D_{(A)}^\m\bar\omega^i+\vO_\m^i\g^{i\m}\bigg]\;. 
\end{split}
\ee
The constant $\gc$ is the gauge coupling and $\a$ is the gauge-fixing
parameter. The field strength and covariant derivatives are
defined in the standard way,
\bseq
\label{YMFD}
\begin{align}
\label{YMF}
&F_{\m\n}^i=\d_\m A^i_\n-\d_\n A^i_\m+f^{ijk}A_\m^j A_\n^k\;,\\
\label{YMD}
&D_{(A)\m}\omega^i=\d_\m\omega^i+f^{ijk}A_\m^j\omega^k\;,
\end{align}
\eseq
and similarly for $D_{(A)\m}\bar\omega^{i}$. The $\B$-covariant derivative
$D_{(\B)\m}\bar\omega^{i}$ is given by an expression analogous to
(\ref{YMgauge}). Clearly, the action (\ref{YMSig}) 
is invariant under gauge
rotations of all fields accompanied by simultaneous gauge
transformations of $A_\m^i$ and $\B^i_\m$: these are precisely the
background-gauge transformations introduced in Sec.~\ref{sec:gauge}.

An important property of the YM theory is
renormalizability. Its key prerequisite are restrictions imposed
on divergences by power counting. The scaling transformations,
\be
\label{YMscale}
x^\mu\mapsto a^{-1}x^\mu~,~~~~
A^i_\m\mapsto a\, A^i_\m\;,
\ee
where $a$ is an arbitrary positive constant,
leave the classical YM action invariant. We will say that
$A^i_\m$ has \emph{scaling} dimension $(+1)$, whereas the dimension of $x^\mu$ is
$(-1)$. The rest of (\ref{YMSig}) will be also invariant if we
simultaneously scale all fields with the following dimensions,
\be
\label{YMdims}
[A^i_\m]=[\B^i_\m]=[\omega^i]=[\bar\omega^{i}]=1~,~~~~~
[b^i]=[\g^{i\m}]=[\z^i]=[\vO^i_\m]=2\;.
\ee 
The textbook analysis of divergent Feynman diagrams
shows that the scaling dimensions of local counterterms needed to
cancel the divergences do not exceed 4. Comparing with the
BRST form (\ref{BRSTSL}) and taking into account that the
generalized BRST operator $\bQ$ increases the scaling dimension by
$1$, we conclude that the dimensions of local operators entering into
the renormalized gauge fermion $\hat\varPsi$ do not
exceed~3. Recalling further that the ghost number of $\hat\varPsi$ is
$(-1)$ we write down the most general expression compatible with
these
requirements,
\be
\label{YMPsi}
\hat\varPsi = \int {\rm d}^4x\,\big(-\hat\g^{i\m}\,U^{i}_{\m}(A,\B)
+\z^i\omega^j V^{ij}\,\big)\;,
\ee 
where $V^{ij}$ are dimensionless constants, 
while $U^{i}_\m$ depends on $A^i_\m$ and 
$\B^i_\m$ at most linearly. We observe that
$\hat\varPsi$ does not depend on $\vO^i_\m$ and is linear in 
$\g^{i\m}$ and $\z^i$. As discussed in Sec.~\ref{sec:2.4}, this
implies that the field redefinition needed to bring the counterterms
into the BRST form is independent of the auxiliary BRST sources, 
see (\ref{reparamrenorm}). 
Positive dimensions of the
YM field and ghosts further constrain this reparameterization to be linear. 

%%%%%%%%%%%%%%%%%%%%%%%%%%%%
\subsection{Higher-derivative relativistic gravity in $(3+1)$ dimensions}
\label{sec:SG}
%%%%%%%%%%%%%%%%%%%%%%%%%%%%

The
fields describing relativistic gravitational theories are identified as follows:
\be
\label{HDfields}
\varphi^a\mapsto g_{\m\n}(x)~,~~~~\varepsilon^\a\mapsto \varepsilon^\m(x)\;,
\ee
where $g_{\m\n}(x)$ is the spacetime metric and 
$\varepsilon^\m(x)$ is a vector field generating infinitesimal
diffeomorphisms. The gauge transformations read,
\be
\label{HDtrans}
\delta_\varepsilon g_{\m\n}=\ve^\l\d_\l g_{\m\n}
+g_{\m\l}\d_\n\ve^\l+g_{\n\l}\d_\m\ve^\l
=\nabla_{(g)\m}\,\ve_\n+\nabla_{(g)\n}\,\ve_\m\;,
\ee
where in the last equality we have lowered the indices using the metric
$g_{\m\n}$ and introduced the covariant derivative $\nabla_{(g)}$
constructed using this metric. The expression (\ref{HDtrans}) implies,
\bseq
\label{HDgens}
\begin{gather}
R^a_{~b\a}\mapsto\delta_\m^\rho\left[\delta_\n^\s
\big(\d_\l\delta(x\!-\!x_1)\big)\,\delta(x\!-\!x_2)
+\delta_\l^\s\,
\delta(x\!-\!x_1)\,\d_\nu\delta(x\!-\!x_2)\right]
+\delta_\n^\rho\delta_\l^\s\,
\delta(x\!-\!x_1)\,\d_\m\delta(x\!-\!x_2)\;,\\
P^a_{~\a}=0~,~~~~~
C^\a_{~\b\g}\mapsto\,\delta^\m_\l\delta(x-x_1)\,\d_\n\delta(x-x_2)
-\delta^\m_\n\,\big(\d_\l\delta(x-x_1)\big)\,\delta(x-x_2)\;.
\end{gather}
\eseq
We focus on the theory in $(3+1)$ dimensions including up to $4$-th
order derivatives of the metric. The  classical action reads 
\be
\label{HDS}
S=\int {\rm d}^4x\sqrt{|g|}\,\bigg[\frac{1}{f_1^2}R_{\m\n}R^{\m\n}
+\frac{1}{f_2^2}R^2-\frac{1}{2\kappa^2}R+\frac{\Lambda}{\kappa^2}\bigg]\;,
\ee
where $|g|$ is the determinant of the metric, $R_{\m\n}$ is the
corresponding Ricci
tensor and $R\equiv R_{\m\n}g^{\m\n}$ is the Ricci scalar; $f_1^2$,
$f_2^2$, $\kappa^2$ and $\Lambda$ are coupling constants. 
The quantum properties of this theory were first analyzed in
\cite{Stelle:1976gc}. The fact that 
the action contains fourth derivatives of the
metric entails well-known problems with the physical
interpretation of the theory \cite{Stelle:1977ry}. 
However, this issue is
irrelevant for our purposes.  

Introducing the background metric $\mathfrak{g}_{\m\n}(x)$ we consider the
gauge fixing function,
\be
\label{HDchi}
\chi^\a\mapsto 
\chi^\m=\mg^{\m\l}\mg^{\n\rho}\Box_{(\mg)}
\nabla_{(\mg)\n}(g_{\l\r}-\mg_{\l\r})\;,
\ee 
where $\nabla_{(\mg)}$ and $\Box_{(\mg)}$ stand for the covariant
derivatives and d'Alembertian constructed from the {\em background
  metric}. 
Introducing the fields of the BRST sector,
\be
\label{HDaux}
\omega^\a\mapsto\omega^\m(x)\,,~~
\bar \omega_\a\mapsto\bar\omega_\m(x)\,,~~
b_\a\mapsto b_\m(x)\,,~~
\g_a\mapsto\g^{\m\n}(x)\,,~~
\z_\a\mapsto\z_\m(x)\,,~~
\vO^a\mapsto\vO_{\m\n}(x)\;,
\ee
and the operator $O^{\a\b}$, 
\be
\label{HDO}
O^{\a\b}\mapsto -\a\frac{\mg^{\m\n}}{\sqrt{|\mg|}}\Box_{(\mg)}
\delta(x-x_1)\;,
\ee
we arrive at the gauge-fixed action,
\be
\label{HDSig}
\begin{split}
\vS_0&=S[g_{\m\n}]+\int {\rm d}^4x\bigg\{b_\m\chi^\m
+\frac{\a}{2}b_\m\frac{\mg^{\m\n}}{\sqrt{|\mg|}}\Box_{(\mg)}b_\n
+\big(\nabla_{(\mg)\n}\bar\omega_\m\big)\mg^{\m\l}\mg^{\n\r}
\Box_{(\mg)}\big(\nabla_{(g)\l}\omega_\r\!+\!
\nabla_{(g)\r}\omega_\l\big)\\
&+\!\g^{\m\n}\big(\nabla_{(g)\m}\omega_\n\!+\!
\nabla_{(g)\n}\omega_\m\big)\!+\!\z_\m\omega^\l\d_\l\omega^\m
\!+\!\vO_{\m\n}\g^{\m\n}
\!+\!\vO_{\m\n}\bar\omega_\l
\bigg[\frac{\delta\chi^\l}{\delta\mg_{\m\n}}
+\frac{\a}{2}\frac{\delta}{\delta\mg_{\m\n}}
\bigg(\frac{\mg^{\l\r}}{\sqrt{|\mg|}}\Box_{(\mg)}\bigg)\,b_\r\bigg]
\bigg\}\,.
\end{split}
\ee
We have not expanded the variational derivatives in the last term as
the corresponding expressions are rather lengthy and not
informative. The background-gauge transformations correspond to
diffeomorphisms,
\be
\label{HDdiff}
x^\m\mapsto x^\m+\ve^\m(x)\;,
\ee 
under which $g_{\m\n}$, $\mg_{\m\n}$, $\omega^\m$, $\vO_{\m\n}$
transform as tensors, whereas $\bar\omega_\m$, $b_\m$, $\g^{\m\n}$,
$\z_\m$ transform as vector/tensor densities. For example,
\bseq
\begin{align}
&\delta_\ve\omega^\m=\ve^\l\d_\l\omega^\m-\omega^\l\d_\l\ve^\m\;,\\
&\delta_\ve\bar\omega_\m=\ve^\l\d_\l\bar\omega_\m+
\bar\omega_\l\d_\m\ve^\l+\bar\omega_\m\d_\l\ve^\l\;,
\end{align}
\eseq
and similarly for the rest of the fields. It is straightforward to see
that this is a symmetry of the action (\ref{HDSig}). 
The fact that $b_\m$ is a covariant vector density explains the
unusual placement of $\sqrt{|\mg|}$ in the denominator of the operator
(\ref{HDO}). 

The four-derivative terms in the classical action (\ref{HDS}) are
invariant under rescaling
\[x^\m\mapsto a^{-1}x^\m\, ,\] 
with the
metric $g_{\m\n}$ kept intact. The same is true for the BRST-exact
part of (\ref{HDSig}) if we assign the following scaling dimensions,
\be
\label{HDdims}
[g_{\m\n}]=[\mg_{\m\n}]=[\omega^\m]=[\bar\omega_\m]=0~,~~~~
[b_\m]=[\vO_{\m\n}]=1~,~~~~
[\g^{\m\n}]=[\z_\m]=3\;.
\ee
As in the case of YM, it can be shown\footnote{The choice of
  gauge (\ref{HDchi}) is important for the argument. It ensures that the
  propagators of the metric perturbations and ghosts
 fall off as the
  fourth power of momentum and as a consequence the degree of
  divergence of Feynman diagrams is consistent with the naive power
  counting.} \cite{Stelle:1976gc}
that the power-counting restricts the scaling dimensions of
counterterms in the Lagrangian to be less than or equal to $4$. This again
constrains the dependence of the gauge fermion on
the auxiliary fields. We observe that the BRST transformations
increase the scaling dimension of all fields\footnote{In this case it is
  due to the presence of derivatives acting on the transformed field,
  rather than the non-zero dimension of ghosts as it happens for
  YM.} by $1$. This implies that $\hat\varPsi$ should contain
local operators of dimensions not higher than $3$. Besides, their
ghost number must be equal to $(-1)$. 
Taking into account the scaling dimensions
(\ref{HDdims}) and the ghost charges (\ref{Qgh}) we obtain the most
general expression,
\be
\label{HDPsi}
\hat\varPsi=\int {\rm d}^4x\,\big(
-\hat\g^{\m\n}\,U_{\mu\nu}(g,\mg)
+\z_\m\omega^\n\, V_\n^\m(g,\mg)\big),
\ee 
where $U_{\m\n}$, $V^\m_\n$ are dimensionless functions of the
quantum and background metric fields that transform covariantly under
background diffeomorphisms. 
We observe that, similarly to YM, $\hat\varPsi$ is linear in 
the BRST sources $\g$ and $\z$.
However, since the scaling dimension of both
metrics is zero, the coefficients in (\ref{HDPsi}) can depend
nonlinearly on  
$g_{\m\n}$ and $\mg_{\m\n}$. This implies that the
field redefinition (\ref{reparamrenorm}) required to restore the BRST
structure of the renormalized 
action is genuinely nonlinear, cf.~\cite{Stelle:1976gc}. 
 
It is worth noting that the original proof of
renormalizability of the theory (\ref{HDS}) presented in
\cite{Stelle:1976gc} is tied to specific gauges 
where the structure of divergences is particularly simple due to some
special features of the action. For more general gauges,
Ref.~\cite{Stelle:1976gc} took the cohomological structure of
divergences as an assumption. Our results provide a proof of this
structure for a general background gauge and, in this respect,
complement the analysis of~\cite{Stelle:1976gc}.

%%%%%%%%%%%%%%%%%%%%%%%%%%%%
\subsection{Projectable Ho\v rava gravity in $(d+1)$ dimensions} 
\label{sec:HG}
%%%%%%%%%%%%%%%%%%%%%%%%%%%%

Consider a $(d+1)$-dimensional spacetime with 
Arnowitt--Deser--Misner (ADM) decomposition of the metric,
\be
\label{ADM}
\di s^2=N^2{\rm d}t^2+g_{ij}({\rm d}{\bf x}^i+N^i {\rm d}t)({\rm d}{\bf x}^j+N^j {\rm d}t)\;,
\ee
where the indices $i,j=1,\dots,d$ denote spatial
directions\footnote{We do not use color YM indices in this
  subsection, so there should be no confusion with the notations of
  Sec.~\ref{sec:YM}.}; they are raised and lowered using the spatial
metric $g_{ij}$. Let us impose the so-called ``projectability'' 
constraint that the lapse $N$ is not
dynamical and fix $N=1$. This constraint is compatible with a subgroup of time-dependent
diffeomorphisms along spatial directions. Thus, the fields and gauge
parameters are identified as follows,
\be
\label{HGfields}
\vf^a\mapsto g_{ij}(t,{\bf x}),~ N^i(t,{\bf x})~,~~~~~~
\ve^\a\mapsto\ve^i(t,{\bf x})\;.
\ee 
The gauge generators and the structure constants are given by the
corresponding reduction of Eqs.~(\ref{HDgens}). The classical action
is taken in the form \cite{Horava:2008ih},
\be
\label{HGS}
S=\frac{1}{2\kappa^2}\int {\rm d}t {\rm d}^d{\bf x}\sqrt{|g|}
\big(K_{ij}K^{ij}-\l K^2+{\cal V}(g_{ij})\big)\;,
\ee
where
\be
\label{HGextr}
K_{ij}=\frac{1}{2}(\dot g_{ij}-\nabla_{(g)i}N_j-\nabla_{(g)j}N_i)
\ee
is the extrinsic curvature on the constant-time slices and $K\equiv
K_{ij}g^{ij}$ is its trace. Here dot denotes derivative with respect
to time and covariant derivatives $\nabla_{(g)}$ are constructed using
the spatial metric $g_{ij}$; $\kappa$ and $\lambda$ are coupling
constants. The potential ${\cal V}$ contains all local terms invariant under
spatial diffeomorphisms that can be constructed from the spatial metric
$g_{ij}$ using no more than $2d$ spatial derivatives; generically, it
is a finite polynomial of the $d$-dimensional Riemann tensor and its
covariant derivatives. 
Clearly, the action (\ref{HGS}) does not possess
Lorentz symmetry. On the other hand, its highest-derivative part is
invariant under anisotropic (Lifshitz) scaling transformations,
\be
\label{HGscale}
{\bf x}\mapsto a^{-1}\,{\bf x}~,~~~~~~
t\mapsto a^{-d}\,t\;,
\ee  
with the scaling dimensions of the fields,
\be
\label{HGdims1}
[g_{ij}]=0~,~~~~~[N^i]=d-1\;.
\ee
Note that different components of the gauge fields (the components of the ADM metric \eqref{ADM} in this case)
have different dimensions which is a common situation in 
theories with Lifshitz
scaling.  

A background-gauge fixing  procedure compatible with the scaling symmetry
(\ref{HGscale}) was constructed in \cite{Barvinsky:2015kil}. We
introduce the background fields $\mg_{ij}(t,{\bf x})$, $\mN^i(t,{\bf x})$ 
and the combinations 
\be
\label{HGfluct}
h_{ij}=g_{ij}-\mg_{ij}~,~~~~~~
n^i=N^i-\mN^i\;.
\ee
Then the gauge-fixing function reads,
\bseq
\be
\label{HGchi}
\chi^\a\mapsto \chi^i=D_tn^i+\frac{\a}{2}O^{ij}\mg^{kl}
\big(\nabla_{(\mg)k}h_{lj}-\l\nabla_{(\mg)j}h_{kl}\big)\;,
\ee
where
\be
\label{HGDn}
D_tn^i=\dot n^i-\mN^k\nabla_{(\mg)k}n^i
+\nabla_{(\mg)k}\mN^i\,n^k\;,
\ee
and the operator $O^{ij}$ has the 
form\footnote{Note that $O^{ij}$ corresponds to the operator denoted
  by ${\cal O}^{-1}_{ij}$ in Ref.~\cite{Barvinsky:2015kil}.},
\be
\label{HGOij}
O^{ij}=(-1)^{d-1}\nabla_{(\mg)}^{k_1}\ldots\nabla_{(\mg)}^{k_{d-2}}
\big(\D_{(\mg)}\mg^{ij}+\upxi \nabla^i_{(\mg)}\nabla^j_{(\mg)}\big)
\nabla_{(\mg)k_{d-2}}\ldots\nabla_{(\mg)k_1}\;.
\ee
\eseq
Here the covariant spatial Laplacian $\D_{(\mg)}$ and all covariant
derivatives $\nabla_{(\mg)}$ are defined using the background metric
$\mg_{ij}$ with their indices raised and lowered using the same
metric; the constants $\a$ and $\upxi$ are gauge parameters. This gauge fixing
term  satisfies all the requirements formulated in
Sec.~\ref{sec:gauge}: it is linear in the difference between the
quantum and background fields, 
and
covariant under simultaneous gauge transformations of these fields. 
 
We now introduce the rest of objects entering in the BRST construction,
\bseq
\label{HGother}
\begin{gather}
\omega^\a\mapsto \omega^i(t,{\bf x})\,,~~
\bar\omega_\a\mapsto \bar\omega_i(t,{\bf x})\,,~~
b_\a\mapsto b_i(t,{\bf x})\,,~~
O^{\a\b}\mapsto\frac{\a}{\sqrt{|\mg|}}O^{ij}\delta(t-t')
\delta({\bf x}-{\bf x}')\,,\\
\g_a\mapsto\big\{\g^{ij}(t,{\bf x}),~\upgamma_i(t,{\bf x})\big\}\,,~~~~~
\z_\a\mapsto \z_i(t,{\bf x})\,,~~~~~
\vO^a\mapsto\big\{\vO_{ij}(t,{\bf x}),~\Omega^i(t,{\bf x})\big\}\;.
\end{gather}
\eseq
The full gauge-fixed action is lengthy and we do not write it 
explicitly. Importantly, with an appropriate
assignment of dimensions to the fields it is invariant under the
scaling transformations (\ref{HGscale}). By inspection of the
gauge-fixing and the Faddeev--Popov ghost terms we find,
\bseq
\label{HGdims*}
\be
\label{HGdims2}
[\omega^i]=[\bar\omega_i]=0~,~~~~~[b_i]=1\;.
\ee 
The background fields inherit the dimensions from their dynamical
counterparts,
\be
\label{HGdims3}
[\mg_{ij}]=0~,~~~~~[\mN^i]=d-1\;.
\ee 
To determine the dimensions of the auxiliary fields $\g^{ij}$,
$\upgamma_i$, $\z_i$ recall that they couple to the BRST variations 
$\bs g_{ij}$, $\bs N^i$, $\bs \omega^i$ respectively. The latter have
the same form as in relativistic gravity and thus contain one spatial
derivative acting on the fields. This yields,
\be
\label{HGdims4}
[\bs g_{ij}]=[\bs \omega^i]=1~,~~~~~[\bs N^i]=d\;.
\ee
Then, the scale invariance of the terms $\g_a\bs\vf^a$, $\z_\a\bs\omega^\a$
in the action
requires
\be
\label{HGdims5}
[\upgamma_i]=d~,~~~~~[\g^{ij}]=[\z_i]=2d-1\;.
\ee 
Finally, the
coupling $\vO^a\g_a$ present in the action fixes
the dimensions of $\vO_{ij}$, $\Omega^i$,
\be
\label{HGdims6}
[\vO_{ij}]=1~,~~~~~[\Omega^i]=d\;.
\ee 
\eseq
The results of \cite{Barvinsky:2015kil} imply that in this theory 
the ultraviolet divergences
consist of local operators with scaling dimensions not higher than
$2d$. 
Thus, all assumptions of Sec.~\ref{sec:2} are satisfied and according
to Sec.~\ref{sec:2.4} the renormalizad action has the form (\ref{BRSTSL}).
As the BRST transformations increase the dimensionality of the fields
by unity, the
renormalized gauge fermion $\hat\varPsi$ appearing in \eqref{BRSTSL} contains operators with
dimensions less or equal to $(2d-1)$. The general expression
satisfying this property and having the ghost number $(-1)$ reads,
\be
\label{HGPsi}
\hat\varPsi=\int {\rm d}t{\rm d}^d{\bf x}\,\big(-\hat\g^{ij}\,U_{ij}(\g,\mg)
-\hat\upgamma_i\, {\rm U}^i(g,N,\mg,\mN)
+\z_i\,\omega^j\, V^i_j(g,\mg)\big)
\;,
\ee  
where $U_{ij}$, $V^i_j$ are dimensionless functions of $g_{ij}$,
$\mg_{ij}$, whereas ${\rm U}^i$ can also linearly depend on $N^i$,
$\mN^i$. Once more we observe that $\hat\varPsi$ is independent of
$\vO_{ij}$, $\Omega^i$ and depends linearly on the rest of BRST
sources.

Establishing the BRST structure of counterterms in the projectable
Ho\v rava gravity together with the results of
Ref.~\cite{Barvinsky:2015kil}  completes the proof of
renormalizability of this theory.

%%%%%%%%%%%%%%%%%%%%%%%%%%%%
\subsection{General relativity  as effective field theory in $(3+1)$ dimensions}
\label{sec:GR}
%%%%%%%%%%%%%%%%%%%%%%%%%%%%

As an example of a non-renormalizable theory we consider Einstein's
general relativity in $(3+1)$ dimensions. The field content and gauge transformations are
the same as in Sec.~\ref{sec:SG}. What differs is the structure of the
classical action which now reads,
\be
\label{GRS}
S=\frac{1}{2\kappa^2}\int {\rm d}^4x\sqrt{|g|}\, (2\Lambda-R+\ldots)\;,
\ee
where dots stand for an infinite sum of various local scalar operators
constructed from the Riemann tensor and its derivatives. They can be
ordered according to the total number of derivatives $n$ they
contain\footnote{Thus, the terms $R_{\m\n}R^{\m\n}$ and $R^2$ contain
  4 derivatives ($n=4$), $R_{\m\n\l\r}R^{\l\r\s\tau}R_{\s\tau}^{~~\;\m\n}$
  contains 6 derivatives ($n=6$), etc.}. 
At each fixed order, the number of possible terms is finite (though it
grows quickly with $n$). In the spirit of effective field theory, the
higher derivative
contributions are treated as corrections to the terms explicitly shown
in (\ref{GRS}). In particular, the graviton 
propagator is determined from the Einstein--Hilbert part and falls
off as $p^{-2}$ at large momenta $p$.

The background-covariant 
gauge-fixing function can be chosen as in Eq.~(\ref{HDchi}). A yet
simpler choice is provided by 
\be
\label{GRchi}
\chi^\a\mapsto \chi^\m=\mg^{\m\l}\mg^{\n\r}\nabla_{(\mg)\n}
(g_{\l\r}-\mg_{\l\r})\;,
\ee
where $\mg_{\m\n}$ is the background metric. The rest of the gauge
fixing construction proceeds in complete analogy with
Sec.~\ref{sec:SG}. In the present case there are no power-counting
arguments constraining the dependence of divergences on auxiliary
fields. 
Still, the proposition formulated in Sec.~\ref{sec:2.4} ensures that
they are compatible with the BRST structure. The 
field renormalization required to recover this structure is expected 
to have the
general form (\ref{reparam}) and involve ghosts and
auxiliary fields in a nonlinear manner.  
 
%%%%%%%%%%%%%%%%%%%%%%%%%%%%
\section{Equations for the effective action}
\label{sec:4}
%%%%%%%%%%%%%%%%%%%%%%%%%%%%

We now derive the equations obeyed by the effective action $\G_L$ defined in \eqref{Gamma} 
corresponding to the generating functional of the form
(\ref{Wgen}). We will omit the loop index $L$ in this section.

As shown in Appendix~\ref{app:STW},
the closure of $\vS$ under the action of the extended BRST
operator, $\bQ\vS=0$, together with the absence of anomalies, implies the 
Slavnov--Taylor
identity for the partition function,
\be
\label{STW}
\bigg[-J_a\frac{\delta}{\delta\g_a}
+\bar\xi_\a\frac{\delta}{\delta\z_\a}
+\xi^\a\frac{\delta}{\delta y^\a}
+\varOmega^a\frac{\delta}{\delta \phi^a}\bigg]W=0\;.
\ee
Whereas the invariance of $\vS$ and $\varPsi$ under background
gauge transformations leads to the Ward identities, 
\be
\label{WardW}
\begin{split}
\bigg[&-J_a R^a_{~b\a}\frac{\delta}{\delta J_b}
+C^\g_{~\b\a}\bar\xi_\g\frac{\delta}{\delta\bar\xi_\b}
-C^\b_{~\g\a}\xi^\g\frac{\delta}{\delta\xi^\b}
-C^{\b}_{~\g\a}y^\g\frac{\delta}{\delta y^\b}\\
&\qquad\qquad+R^a_{~\a}(\phi)\frac{\delta}{\delta \phi^a}
-\g_bR^b_{~a\a}\frac{\delta}{\delta\g_a}
+C^\b_{~\g\a}\z_\b\frac{\delta}{\delta\z_\g}
+R^a_{~b\a}\varOmega^b\frac{\delta}{\delta\varOmega^a}\bigg]W=0\;.
\end{split}
\ee
Besides, the equations of motion
for the Lagrange multiplier $b_\a$ imply,
\be
\label{beqW}
\bigg[\chi^\a_a\frac{\delta}{\delta J_a}
-O^{\a\b}\frac{\delta}{\delta y^\b}
-\frac{\varOmega^a}{2}\frac{\delta O^{\a\b}}{\delta
  \phi^a}\frac{\delta}{\delta\xi^\b}+y^\a\bigg]W=0\;.
\ee
Let us stress that the derivation of Eqs.~(\ref{WardW}), (\ref{beqW})
essentially 
relies on
the property that the gauge
generators and the gauge-fixing condition are linear in the quantum field.  

Turning to the effective action, we
use the relations (\ref{avfields}), (\ref{sources})
and rewrite the identities (\ref{STW}), (\ref{WardW}) and (\ref{beqW})
in the following
form\footnote{Recall that we omit averaging symbols on the mean fields.},
\bseq
\label{identsG}
\begin{align}
&\frac{\delta\G}{\delta\g_a}\frac{\delta\G}{\delta\varphi^a}
+\frac{\delta\G}{\delta\z_\a}\frac{\delta\G}{\delta \omega^\a}
+b_\a\frac{\delta\G}{\delta\bar\omega_\a} +
\varOmega^a\frac{\delta\G}{\delta \phi^a}=0\;,
\label{STG}\\
&R^a_{~\a}(\varphi)\frac{\delta\G}{\delta \varphi^a}
-C^\g_{~\b\a}\omega^\b\frac{\delta\G}{\delta\omega^\g}
+\bar\omega_\b C^\b_{~\g\a}\frac{\delta\G}{\delta\bar\omega_\g}
+b_\b C^\b_{~\g\a}\frac{\delta\G}{\delta b_\g}\notag\\
&\qquad\qquad+R^a_{~\a}(\phi)\frac{\delta\G}{\delta \phi^a}
-\g_bR^b_{~a\a}\frac{\delta\G}{\delta\g_a}
+\z_\b C^\b_{~\g\a}\frac{\delta\G}{\delta\z_\g}
+R^a_{~b\a}\varOmega^b\frac{\delta\G}{\delta\varOmega^a}=0\;,
\label{WardG}\\
&\chi^\a_a(\varphi^a-\phi^a)-O^{\a\b} b_\b
-\frac{\varOmega^a}{2}\frac{\delta O^{\a\b}}{\delta \phi^a}\bar\omega_\b
-\frac{\delta\G}{\delta b_\a}=0\;.
\label{beqG}
\end{align}
\eseq
It is convenient to consider a reduced effective action $\hat\G$
obtained from $\G$ by subtracting the gauge-fixing term and
its
derivatives with respect to the background fields,
\be
\label{redG}
\hat\G=\G- b_\a\bigg(\chi^\a_a(\varphi-\phi)^a
-\frac{1}{2}O^{\a\b} b_\b\bigg)
-\varOmega^a\bar\omega_\a\bigg(\frac{\delta\chi^\a_b}{\delta \phi^a}
(\varphi-\phi)^b
-\chi^\a_a-\frac{1}{2}\frac{\delta O^{\a\b}}{\delta \phi^a}b_\b\bigg)\;.
\ee
Substituting this expression  into (\ref{beqG}) yields that $\hat\G$ is independent of
$b_\a$,
\be
\label{bindep}
\frac{\delta\hat\G}{\delta b_\a}=0\;.
\ee
Then the identity (\ref{STG}) splits into two equations,
\bseq
\label{STGs}
\begin{align}
\label{STG1}
&\chi^\a_a\frac{\delta\hat\G}{\delta\g_a}
+\frac{\delta\hat\G}{\delta\bar\omega_\a}=0\;,\\
\label{STG2}
&\frac{\delta\hat\G}{\delta\g_a}\frac{\delta\hat\G}{\delta\varphi^a}
+\frac{\delta\hat\G}{\delta\z_\a}\frac{\delta\hat\G}{\delta\omega^\a}
+\varOmega^a\bigg(\frac{\delta\hat\G}{\delta \phi^a}
+\bar\omega_\a\frac{\delta\chi^\a_b}{\delta \phi^a}
\frac{\delta\hat\G}{\delta \g_b}\bigg)
=0\;.
\end{align}
\eseq
The first one implies that $\hat\G$ depends on the antighost only
through the combination (\ref{gahat}), 
so that
\be
\hat\G=\hat\G[\,\varphi,\omega,\phi,\hat\g,\z,\varOmega\,]\;.
\ee
Next, we use the relation
\be
\label{phiders}
\frac{\delta}{\delta\phi^a}\bigg|_{\hat\g}=
\frac{\delta}{\delta\phi^a}\bigg|_{\g}
+\bar\omega_\a\frac{\delta\chi^\a_b}{\delta\phi^a}\frac{\delta}{\delta\g^b}\;,
\ee
where the index on the right of the vertical line means that the
$\phi$-derivative is taken at fixed $\hat\g$ or $\g$.
Consequently, Eq.~(\ref{STG2}) takes the form,
\bseq
\label{identshatG}
\be
\label{SThatG}
\frac{\delta\hat\G}{\delta\hat\g_a}\frac{\delta\hat\G}{\delta\varphi^a}
+\frac{\delta\hat\G}{\delta\z_\a}\frac{\delta\hat\G}{\delta \omega^\a}
+\varOmega^a\frac{\delta\hat\G}{\delta \phi^a}=0\;.
\ee
The Ward identities (\ref{WardG}) also simplify to,
\be
\label{WardhatG}
R^a_{~\a}(\varphi)\frac{\delta\hat\G}{\delta \varphi^a}
-C^\g_{~\b\a}\omega^\b\frac{\delta\hat\G}{\delta\omega^\g}
+R^a_{~\a}(\phi)\frac{\delta\hat\G}{\delta \phi^a}
-\hat\g_bR^b_{~a\a}\frac{\delta\hat\G}{\delta\hat\g_a}
+\z_\b C^\b_{~\g\a}\frac{\delta\hat\G}{\delta\z_\g}
+R^a_{~b\a}\varOmega^b\frac{\delta\hat\G}{\delta\varOmega^a}=0.
\ee
\eseq
Finally, $\hat\G$ has zero ghost number, i.e. it is invariant under
phase rotations of the fields $\omega$, $\hat\g$, $\z$, $\vO$ with
charges (\ref{Qgh}). Together with Eqs.~(\ref{identshatG}) this
will be used in the next section to constrain the structure of
ultraviolet divergences.

Clearly, the identities (\ref{identshatG})
are satisfied by the {\em reduced} tree-level action
$\hat\varSigma_0$, which is related to (\ref{Sig}) by a formula
analogous to (\ref{redG}). Explicitly, we have,
\be
\label{hatS0}
\hat\varSigma_0=S[\,\varphi\,]+\hat\g_aR^a_{~\a}(\varphi)\,\omega^\a
+\frac{1}{2}\z_\a C^\a_{~\b\g}\omega^\b\omega^\g\;.
\ee
Note that $\hat\varSigma_0$ does not have any explicit 
dependence\footnote{Only an implicit dependence of
$\hat\varSigma_0$ on $\phi^a$ through
the combination (\ref{gahat}) remains.} on 
$\varOmega^a$ and $\phi^a$. Consequently, the
last term in (\ref{SThatG}) and the third term in (\ref{WardhatG}) are
absent in the corresponding identities for~$\hat\varSigma_0$.

%%%%%%%%%%%%%%%%%%%%%%%%%%%%
\section{Structure of divergences}
\label{sec:5}
%%%%%%%%%%%%%%%%%%%%%%%%%%%%

We return to the renormalization procedure. Let us assume that at the
order of $(L-1)$ loops we have already shown that the renormalized
generating functional $W_{L-1}$ has the form
(\ref{Wgen})---(\ref{reparam1}).
The first
divergence of the effective action $\G_{L-1}$ appears at order
$\hbar^L$ and is local, see Eqs.~(\ref{GL-1}), (\ref{localdiv}). The
standard procedure prescribes to subtract it from $\vS_{L-1}$ in order to
obtain the action renormalized at $L$ loops. Our task is to work out
the structure of this divergence. To avoid cluttered notations
we will omit the indices related to the induction step
and will denote the relevant divergent part $\G_{L,\infty}$ simply as
$\G_\infty$. 

First, we observe that the transformation (\ref{redG}) involves only
finite quantities, so that the divergent parts of $\G$ and $\hat\G$
coincide,
\be
\label{hatGinf}
\G_\infty=\hat\G_\infty[\vf,\omega,\phi,\hat\g,\z,\vO]\,.
\ee
Due to the linearity of the Ward identities (\ref{WardhatG}), they are
obeyed separately by each term in the expansion of $\hat\G$ in
$\hbar$; in particular, they hold for the divergent part
$\hat\G_\infty$.  
Next, we consider Eq.~(\ref{SThatG}). 
The first divergent contribution into it appears at
the order $\hbar^L$. Equation~(\ref{SThatG}) at this order then implies 
\be
\label{SThatn}
\bQ_+\hat\G_\infty=0\;,
\ee
where we have introduced an operator $\bQ_+$ that acts on a functional
$X$ of the fields $\vf$, $\omega$, $\phi$, $\hat\g$, $\z$ , $\vO$ as follows,
\be
\label{antibracket}
\begin{split}
\bQ_+X
&=\frac{\delta\hat\varSigma_0}{\delta\hat\gamma_a}
\frac{\delta X}{\delta \varphi^a}
+\frac{\delta\hat\varSigma_0}{\delta \vf^a}
\frac{\delta X}{\delta\hat\g_a}
+\frac{\delta\hat\varSigma_0}{\delta\z_\a}
\frac{\delta X}{\delta\omega^\a}
+\frac{\delta\hat\varSigma_0}{\delta\omega^\a}
\frac{\delta X}{\delta\z_\a}
+\vO^a\frac{\delta X}{\delta\phi^a}\\
&\equiv (\hat\vS_0,X)+\vO^a\frac{\delta X}{\delta\phi^a}
\;.
\end{split}
\ee
Here $\hat\vS_0$ is the reduced tree-level action (\ref{hatS0}) and in
the second line we defined the {\em antibracket} $(\hat\vS_0,X)$.
A straightforward calculation using the structural relations
(\ref{compat*}), (\ref{Jacobi}) shows that the latter is nilpotent,
\bseq
\label{antiprops}
\be
\label{nilbrack}
(\hat\vS_0,(\hat\vS_0,X))=0\;,
\ee
and anticommutes with the operator $\vO\, \delta/\delta\phi$,
\be
\label{brackantiO}
(\hat\vS_0,\vO^a \frac{\delta X}{\delta\phi^a})
=-\vO^a \frac{\delta }{\delta\phi^a}(\hat\vS_0,X)\;.
\ee
\eseq
The properties (\ref{antiprops}) imply nilpotency of $\bQ_+$.
Note that using the explicit form of $\vS_0$ and the BRST transformations
(\ref{BRST*}),  $\bQ_+$ can be written as
\be
\label{Q+Q}
\bQ_+X=(\bs\vf^a)\frac{\delta X}{\delta\vf^a}+
(\bs\omega^\a)\frac{\delta X}{\delta\omega^\a}
+\vO^a\frac{\delta X}{\delta\phi^a}\bigg|_{\hat\g}
+\frac{\delta\hat\varSigma_0}{\delta \vf^a}
\frac{\delta X}{\delta\hat\g_a}
+\frac{\delta\hat\varSigma_0}{\delta\omega^\a}
\frac{\delta X}{\delta\z_\a}\;.
\ee  
The first three terms here resemble the action of the operator $\bQ$
introduced in Sec.~\ref{sec:gauge}. However, there are a few
differences. $\bQ$ is defined on functionals of all quantum fields
$\vf,\omega,\bar\omega,b$
and
external backgrounds $\phi,\g,\z,\vO$. On the other hand, $\bQ_+$ acts
on functionals that are
restricted to the minimal sector of quantum fields
$\vf,\omega$ and,
instead of $\g$, depend on the combination $\hat\g$ (see
(\ref{gahat})) treated as a free variable. 

We now use Eq.~(\ref{SThatn}) to determine the dependence of
$\hat\G_\infty$ on the background fields $\phi^a$.

%%%%%%%%%%%%%%%%%%%%%%%%%%%%
\subsection{Separating the background field dependence}
\label{sec:5.1}
%%%%%%%%%%%%%%%%%%%%%%%%%%%%

We expand $\hat\G_\infty$ in powers of the auxiliary source $\vO$,
\be
\label{Omegaexp}
\hat\G_\infty=\sum_k\hat\G_{\infty,\{k\}}~,~~~~
\hat\G_{\infty,\{k\}}=\vO^{a_1}\ldots\vO^{a_k}
\hat\G_{\infty,\{k\},[a_1,\ldots,a_k]}[\vf,\omega,\phi,\hat\g,\z]\;.
\ee
We assume that this sum is finite, $k\leq K$, which will be
justified shortly. Substituting (\ref{Omegaexp}) into (\ref{SThatn}) we
obtain 
\bseq
\begin{align}
\label{kcoh1}
&\vO^a\frac{\delta\hat\G_{\infty,\{K\}}}{\delta\phi^a}=0\;,\\
\label{kcoh2}
&\vO^a\frac{\delta\hat\G_{\infty,\{k\}}}{\delta\phi^a}
+(\hat\vS_0,\hat\G_{\infty,\{k+1\}})=0~,~~~~0\leq k\leq K-1\;.
\end{align}
\eseq
As shown in Appendix~\ref{app:coh}, the cohomology of the
operator $\vO\delta/\delta\phi$ on the space of 
local functionals vanishing at $\vO=0$ is
trivial. In other words, Eq.~(\ref{kcoh1}) implies that
$\hat\G_{\infty,\{K\}}$ is represented as
\be
\label{GKsol}
\hat\G_{\infty,\{K\}}=\vO^a\frac{\delta}{\delta\phi^a}\varUpsilon_{\{K-1\}}\;,
\ee 
where $\varUpsilon_{\{K-1\}}$ is a local functional of ghost number
$(-1)$ invariant under background-gauge transformations. Inserting
this representation into (\ref{kcoh2}) for $k=K-1$ yields
\be
\label{kcoh3}
\vO^a\frac{\delta}{\delta\phi^a}\big(\hat\G_{\infty,\{K-1\}}
-(\hat\vS_0,\varUpsilon_{\{K-1\}})\big)=0\;,
\ee
where we have used the property (\ref{brackantiO}). Again, this
implies
\be
\label{GK-1sol}
\hat\G_{\infty,\{K-1\}}=(\hat\vS_0,\varUpsilon_{\{K-1\}})
+\vO^a\frac{\delta}{\delta\phi^a}\varUpsilon_{\{K-2\}}\;.
\ee 
By continuing this reasoning and using the properties
(\ref{antiprops}) we obtain a representation of the type
(\ref{GK-1sol}) for all $\hat\G_{\infty,\{k\}}$, $1\leq k\leq
K-1$. For $k=0$ an additional contribution appears,
\be
\label{G0sol}
\hat\G_{\infty,\{0\}}=(\hat\vS_0,\varUpsilon_{\{0\}})
+\mbox{\boldmath $\varGamma$}\;,
\ee
where $\bG[\vf,\omega,\hat\g,\z]$ is
independent of $\vO$ and the background field $\phi$. Collecting all
contributions together we arrive at
\begin{align}
\hat\G_\infty&=\bG[\vf,\omega,\hat\g,\z]
+\sum_{k=0}^{K-1}(\hat\vS_0,\varUpsilon_{\{k\}})
+\sum_{k=1}^K\vO^a\frac{\delta}{\delta\phi^a}\varUpsilon_{\{k-1\}}\notag\\
&=\bG[\vf,\omega,\hat\g,\z]
+\bQ_+\varUpsilon\;,
\label{Gphisol}
\end{align}
where in the second line we have defined
\be
\label{Usptot}
\varUpsilon[\vf,\omega,\phi,\hat\g,\z,\vO]=\sum_{k=0}^{K-1}\varUpsilon_{\{k\}}\;.
\ee

We can now appreciate the power of the background-field approach.
The pieces dependent on the background fields
have separated into a $\bQ_+$-exact
contribution 
leaving behind the
part $\bG$ that depends only on the quantum fields.  
The original invariance
under background-gauge transformations implies that $\bG$ is
gauge-invariant on its own.
More precisely, we write
\be
\label{calS1}
\bG=\mbox{\boldmath $S$}[\vf]+\varLambda[\vf,\omega,\hat\g,\z]\;, 
\ee
where $\varLambda$ vanishes at $\omega=0$. The ghost-independent part
$\mbox{\boldmath $S$}[\vf]$ cannot depend on $\hat\g$ or $\z$ as the
latter have negative ghost charges (see (\ref{Qgh})), whereas the ghost
number of $\bG$ is zero. Then, due to the Ward identities
(\ref{WardhatG}), the local
functional $\mbox{\boldmath $S$}[\vf]$ satisfies
\be
\label{boldSgt}
\frac{\delta\mbox{\boldmath $S$}}{\delta\vf^a}R^a_{~\a}(\vf)=0\;.
\ee
We will see in Sec.~\ref{sec:5.3} that 
in the subtraction procedure it combines with the classical action
$S[\vf]$ and corresponds to the renormalization of the couplings in the
classical gauge invariant Lagrangian. 
The rest of the terms in (\ref{Gphisol}), (\ref{calS1}) generates
a renormalization of the gauge fermion and the corresponding field
redefinition. \\

We still have to justify the assumption that the sum (\ref{Omegaexp})
can be truncated at finite $k$. We do it using the notion of
derivative expansion.
Being local, the functional $\hat\G_\infty$ is a spacetime integral of
a Lagrangian which can be written as a series of terms, each
of them containing  a finite number of derivatives. Let us introduce a
formal book-keeping parameter $l_*$ of dimension of length 
counting the number of derivatives in
a given term,  and convert the derivative expansion  into a
Taylor series\footnote{In theories with Lifshitz scaling it would be
  natural to assign different weights to 
 derivatives along different
  spacetime directions,
  cf. Sec.~\ref{sec:HG}. However, the argument presented below does not depend
  on whether one introduces such weighting or not, so for simplicity we treat
  all derivatives on equal footing.} in
$l_*$. We denote by $\hat\G_\infty^{N}$ the part of $\hat\G_\infty$
containing all terms of order $l_*^n$, $n\leq N$, i.e. all terms with up
to $N$ derivatives. Now, $\vO$ is an anticommuting local
field. With a finite number of derivatives at disposal, one can
construct only a finite number of local operators out of it. Therefore
$\hat\G_\infty^{N}$ is a {\em finite} polynomial\footnote{Note that
its highest power is not directly related to $N$ and can depend on the
specifics of the theory such as number of internal indices and spacetime
dimensions, power-counting considerations, etc. The only property which
is important for us here is that this power is finite.} in $\vO$. 

Next, we observe that $\hat\vS_0$ is also a local functional and hence
contains derivatives in non-negative powers. Thus, it is represented as
a series with non-negative powers of $l_*$, so that the antibracket
$(\hat\vS_0,...)$ acting on a given operator cannot decrease its order
in $l_*$. Besides, the operator
$\vO\delta/\delta\phi$ does not contain $l_*$ at
all. We conclude that $\hat\G_\infty^{N}$ satisfies
Eq.~(\ref{SThatn}), up to corrections of order $l_*^{N+1}$,
\be
\label{SThatGN}
\bQ_+\hat\G_\infty^{N}={\cal O}(l_*^{N+1})\;.
\ee
Splitting $\hat\G_\infty^{N}$ into monomials in $\vO$ one can
repeat the derivation leading to (\ref{Gphisol}), up to corrections of
order ${\cal O}(l_*^{N+1})$ on the r.h.s. As this representation holds
for any $N$, we can send the latter to infinity\footnote{In
  renormalizable theories with finite number of coupling constants the
derivative expansion usually terminates at a finite order in $N$.} 
and recover
(\ref{Gphisol}) for the full divergent part 
$\hat\G_\infty$ without any corrections.

%%%%%%%%%%%%%%%%%%%%%%%%%%%%
\subsection{Ghost-dependent contribution}
\label{sec:5.2}
%%%%%%%%%%%%%%%%%%%%%%%%%%%%

It remains to fix the structure of the term $\vLam$ in
(\ref{calS1}). It satisfies the equation,
\be
\label{STtilden}
(\hat\vS_0,\vLam)=0\;.
\ee
Using the explicit form of the reduced tree-level action (\ref{hatS0})
and the definition of the antibracket, we obtain
\be
\label{STtilden1}
\begin{split}
R^a_{~\a}(\varphi)\,\omega^\a
\frac{\delta\vLam}{\delta \varphi^a}
+\bigg(\frac{\delta S}{\delta \varphi^a}
&+\hat\g_b R^b_{~a\a}\omega^\a\bigg)
\frac{\delta\vLam}{\delta \hat\g_a}
+\frac{1}{2}C^\a_{~\b\g}\omega^\b\omega^\g
\frac{\delta\vLam}{\delta\omega^\a}\\
&\quad+\Big(-\hat\g_a R^a_{~\a}(\varphi)
+\z_\b C^\b_{~\a\g}\omega^\g\Big)
\frac{\delta\vLam}{\delta\z_\a}=0\;.
\end{split}
\ee
Besides, the invariance of $\vLam$ with respect to background-gauge
transformations implies the Ward identities (cf. (\ref{WardhatG})),
\be
\label{Wtilden1}
R^a_{~\a}(\varphi)
\frac{\delta\vLam}{\delta \varphi^a}
-C^\g_{~\b\a}\omega^\b \frac{\delta\vLam}{\delta\omega^\g}
-\hat\g_b R^b_{~a\a}\frac{\delta\vLam}{\delta\hat\g_a}
+\z_\b C^\b_{~\g\a}\frac{\delta\vLam}{\delta\z_\g}=0\;.
\ee
Multiplying the latter expression by $\omega^\a$ and subtracting it from
(\ref{STtilden1}), we arrive at the equation
\be
\label{STtilden2}
(\bq_0+\bq_1)\,\vLam=0\;,
\ee
where the operators $\bq_{0,1}$ are defined as
\bseq
\label{q01} 
\begin{align}
\label{q0}
&\bq_0\,\vLam=
\frac{\delta S}{\delta \varphi^a}
\frac{\delta\vLam}{\delta\hat\g_a}
-\hat\g_a R^a_{~\a}(\varphi)
\frac{\delta\vLam}{\delta\z_\a}\;,\\
\label{q1}
&\bq_1\,\vLam=
-\frac{1}{2}C^\g_{~\a\b}\omega^\a\omega^\b
\frac{\delta\vLam}{\delta\omega^\g}\;.
\end{align}
\eseq
Both operators are nilpotent and anticommute with each other,
\be
\label{q01props}
(\bq_0)^2=(\bq_1)^2=\bq_0\bq_1+\bq_1\bq_0=0\;.
\ee
The operator $\bq_0$ is known in the mathematical literature as
Koszul--Tate differential \cite{Henneaux:1990rx}.

Let us expand $\vLam$ in powers of the ghost fields $\omega^\a$,
\be
\label{Lamexp}
\vLam=\sum_{k=1}^\infty\vLam^{\{k\}}~,~~~~
\vLam^{\{k\}}=\omega^{\a_1}\ldots\omega^{\a_k}
\vLam^{\{k\}}_{[\a_1,\ldots,\a_k]}[\vf,\hat\g,\z]\;.
\ee
Note that the sum starts at $k=1$ as, by definition, $\vLam$
vanishes at $\omega=0$. The conservation of the ghost number and the ghost
charges (\ref{Qgh}) imply that each term $\vLam^{\{k\}}$ in the
expansion is a finite
polynomial in $\hat\g$ and $\z$ that vanishes at
$\hat\g=\zeta=0$. Thus $\vLam^{\{k\}}$ satisfies,
\be
\label{Lamzero}
\vLam^{\{k\}}\big|_{\omega=0}=\vLam^{\{k\}}\big|_{\hat\g=\z=0}=0\;.
\ee
We now substitute (\ref{Lamexp}) into (\ref{STtilden2}) and obtain a
chain of equations,
\bseq
\begin{align}
\label{Lam1eq}
&\bq_0\vLam^{\{1\}}=0\;,\\
\label{Lamkeq}
&\bq_0\vLam^{\{k\}}+\bq_1\vLam^{\{k-1\}}=0\;,~~~~k\geq 2\;.
\end{align}
\eseq  
The Koszul--Tate differential $\bq_0$ has trivial cohomology on 
functionals satisfying (\ref{Lamzero}) if the gauge algebra obeys the
conditions \ref{(i)}, \ref{(ii)} from
Sec.~\ref{sec:2.1} \cite{BV3}:
\be
\label{cohstat}
\bq_0 X=0~,~~X\big|_{\omega=0}=X\big|_{\hat\g=\z=0}=0~~~\Longrightarrow~~~
X=\bq_0 Y\;.
\ee
Moreover, under natural assumptions about the regularity of
the equations of motion, the functional $Y$ can be chosen to be local 
\cite{Henneaux:1990rx,Vandoren:1993bw}, provided 
$X$ itself is local. Finally, one can show 
along the lines of \cite{Vandoren:1993bw}
that there exists a choice of $Y$ 
which inherits all linearly realized symmetries 
commuting with $\bq_0$. In particular, we can take
$Y[\vf,\omega,\hat\g,\z]$ to be
invariant under background-gauge transformations if so is $X$. 

Thus we
write,
\be
\label{Lam1sol}
\vLam^{\{1\}}=\bq_0\varXi^{\{1\}}\;,
\ee  
where $\varXi^{\{1\}}$ is local and background-gauge invariant.
Substituting this into Eq.~(\ref{Lamkeq}) for $k=2$ and interchanging
the order of $\bq_0$ and $\bq_1$ we obtain,
\be
\label{Lam2eqnew}
\bq_0\big(\vLam^{\{2\}}-\bq_1\varXi^{\{1\}}\big)=0\;,
\ee
whence
\be
\label{Lam2sol}
\vLam^{\{2\}}=\bq_1\varXi^{\{1\}}+\bq_0\varXi^{\{2\}}\;.
\ee
Continuing by induction, we obtain analogous representations for all
$\vLam^{\{k\}}$. Collected together they give,
\be
\label{Lamsol}
\vLam=(\bq_0+\bq_1)\varXi~,~~~~\varXi=\sum_{k=1}^\infty\varXi^{\{k\}}\;.
\ee  
To make the last step, we notice that $\varXi$, due its invariance
under background-gauge
transformations,
obeys
a Ward identity analogous to (\ref{Wtilden1}). 
Combining this with (\ref{Lamsol})
we get,
\be
\label{Lamsolfin}
\vLam=(\hat\vS_0,\varXi)\;.
\ee 
This is our final expression for $\vLam$.\\

Putting together the contributions (\ref{Gphisol}), (\ref{calS1}),
(\ref{Lamsolfin}) and reintroducing  the loop index $L$, we obtain the
desired form of the $L$-loop divergence
\be
\label{Gdiv0}
\G_{L,\infty}=\mbox{\boldmath $S$}_L[\vf]+\bQ_+\bUp_L\;,
\ee
where $\bUp_L=\varUpsilon_L+\varXi_L$ and we have used that $\varXi_L$
is independent of $\phi$.

%%%%%%%%%%%%%%%%%%%%%%%%%%%%
\subsection{Subtraction and field redefinition}
\label{sec:5.3}
%%%%%%%%%%%%%%%%%%%%%%%%%%%%

We now define the $L$-th order renormalized action
as\footnote{Strictly speaking, according to the standard scheme one should
take $\phi^a-\delta\varPsi_{L-1}/\delta\g_a$ and 
$\delta\varPsi_{L-1}/\delta\z_\a$ instead of $\vf^a$ and $\omega^\a$ as
arguments of $\G_{L,\infty}$. However, due to the representation
(\ref{Psihattree}) valid for $\varPsi_{L-1}$, the difference produced
by this replacement is of higher order in $\hbar$. It is included in
the ${\cal O}(\hbar^{L+1})$ term in (\ref{SigL0}).} (compare with \eqref{SigmaL}),
\be
\label{SigL0}
\vS_L[\vf,\omega,\bar\omega,b,\phi,\g,\z,\vO]
=\vS_{L-1}-\hbar^L\G_{L,\infty}[\vf,\omega,\phi,\hat\g,\z,\vO]
+{\cal O}(\hbar^{L+1})\;,
\ee
where the last term on the r.h.s. stands for {\em local} operators
multiplied by at least $\hbar^{L+1}$ that will be specified
shortly. The presence of these operators does not spoil the key property
of the subtraction prescription, namely  that it removes all subdivergences at
$(L+1)$-loop order. Thus, according to the assumption stated in
Sec.~\ref{sec:2.3}, the 
$(L+1)$-loop divergence will be local.

We now show that $\vS_L$ can be brought to the form (\ref{BRSTSL})
by a reparameterization of the fields $\vf$, $\omega$. Substituting the
expression (\ref{Gdiv0}) in \eqref{SigL0} and expanding explicitly the operator $\bQ_+$ 
we obtain,
\be
\label{SigL1}
\begin{split}
\varSigma_L
=\varSigma_0&+\sum_{l=1}^{L-1}\hbar^l\vS_{l}^C-\hbar^L\mbox{\boldmath $S$}_L
-\hbar^L\frac{\delta\bUp_L}{\delta\hat\g_a}\frac{\delta\hat\vS_0}{\delta\vf^a}
+\hbar^L\frac{\delta\bUp_L}{\delta\z_\a}\frac{\delta\hat\vS_0}{\delta\omega^\a}
\\
&-\hbar^L\frac{\delta\hat\vS_0}{\delta\hat\g_a}\frac{\delta\bUp_L}{\delta\vf^a}
-\hbar^L\frac{\delta\hat\vS_0}{\delta\z_\a}\frac{\delta\bUp_L}{\delta\omega^\a}
-\hbar^L\vO^a\frac{\delta\bUp_L}{\delta\phi^a}\bigg|_{\hat\g}
+{\cal O}(\hbar^{L+1})\;.
\end{split}
\ee
As before, the index $\hat\g$ on the partial derivative
with respect to the background field in the last significant term
emphasizes that it is taken at
fixed $\hat\g$. The first two terms in the last line have the form,
\be
\label{BRSTpart}
-\hbar^L\bs\vf^a\,\frac{\delta\bUp_L}{\delta\vf^a}
-\hbar^L \bs\omega^\a\,\frac{\delta\bUp_L}{\delta\omega^\a}\;.
\ee
This suggests to define the $L$-th order gauge fermion,
\bseq
\be
\label{PsiL}
\varPsi_L=\varPsi_{L-1}-\hbar^L\bUp_L\;,
\ee  
and the $L$-th order counterterm
\be
\label{SigCL}
\vS_L^C=-\mbox{\boldmath $S$}_L[\vf]-\bs\bUp_L
-\vO^a\frac{\delta\bUp_L}{\delta\phi^a}\bigg|_\g
=-\mbox{\boldmath $S$}_L[\vf]-\bQ\,\bUp_L\;.
\ee
\eseq
To proceed, we notice that the expressions (\ref{Sig}) and (\ref{hatS0}) 
imply
\bseq
\be
\frac{\delta\vS_0}{\delta\vf^a}=
\frac{\delta\hat\vS_0}{\delta\vf^a}
+b_\a\chi^\a_a+\vO^b\bar\omega_\a\frac{\delta\chi_a^\a}{\delta\phi^b}\;.
\ee
Further, as a consequence of the definition (\ref{gahat}) we have 
\be
\bs \bUp_L=\bs\vf^a\,\frac{\delta\bUp_L}{\delta\vf^a}
+ \bs\omega^\a\,\frac{\delta\bUp_L}{\delta\omega^\a}
-b_\a\chi^\a_a\,\frac{\delta\bUp_L}{\delta\hat\g_a}\;.
\ee
\eseq
Finally, the $\phi$-derivatives at fixed $\g$ and $\hat\g$ are related
by (\ref{phiders}). Collecting all the previous expressions together, we find that  
Eq.~(\ref{SigL1}) simplifies to
\be
\label{SigL2}
\vS_L=\vS_0+\sum_{l=1}^{L}\hbar^l\vS_l^C
-\hbar^L\frac{\delta\bUp_L}{\delta\g_a}\frac{\delta\vS_0}{\delta\vf^a}
+\hbar^L\frac{\delta\bUp_L}{\delta\z_\a}\frac{\delta\vS_0}{\delta\omega^\a}
+{\cal O}(\hbar^{L+1})\;.
\ee
The first two terms already have the desired BRST form (\ref{BRSTSL}),
\be
\label{good}
\vS_0+\sum_{l=1}^{L}\hbar^l\vS_l^C=S[\vf]-\sum_{l=1}^{L}\hbar^l 
\mbox{\boldmath $S$}_L[\vf]+\bQ\,\varPsi_L\;.
\ee
The
remaining contributions are absorbed by a field redefinition, as we
now demonstrate. First we perform the change of variables
$\vf,\omega\mapsto\vf',\omega'$ 
given by
\bseq
\label{redef}
\begin{align}
\label{redefvf}
&\vf^a=\vf'^a
+\hbar^L\frac{\delta\bUp_L}{\delta\g_a}(\vf',\omega',\ldots)
+{\cal O}(\hbar^{L+1})\;,\\
\label{redefomega}
&\omega^\a=\omega'^\a
-\hbar^L\frac{\delta\bUp_L}{\delta\z_\a}(\vf',\omega',\ldots)
+{\cal O}(\hbar^{L+1})\;,
\end{align}
\eseq
where we again allow for possible local contributions of higher order
in $\hbar$. Next, we Taylor expands all quantities in the differences
$(\vf-\vf')$, $(\omega-\omega')$. Then, the third and fourth terms in
(\ref{SigL2}) are cancelled by the linear contribution in the series
for $\vS_0$. Other terms generated by the expansion are of higher
powers in $\hbar$. Notice that they are local. Thus, by properly
adjusting the ${\cal O}(\hbar^{L+1})$ contribution in (\ref{SigL2})
they can be cancelled as well.  

To complete the argument we need to verify that the operators coupled to
sources in the path integral have the right form (\ref{reparam1}) in terms
of the new variables. This is done through the following chain of
relations,
\be
\label{coupling}
\begin{split}
\tilde\vf_{L-1}^a(\vf,\omega,\ldots)-\phi^a
&=
-\frac{\delta\varPsi_{L-1}}{\delta\g_a}(\vf,\omega,\ldots)\\
&=\vf^a-\phi^a+\sum_{l=1}^{L-1}\hbar^l\,
\frac{\delta\bUp_l}{\delta\g_a}(\vf,\omega,\ldots)\\
&=\vf'^a-\phi^a+\sum_{l=1}^{L-1}\hbar^l\,
\frac{\delta\bUp_l}{\delta\g_a}(\vf,\omega,\ldots)
+\hbar^L\,\frac{\delta\bUp_L}{\delta\g_a}(\vf',\omega',\ldots)
+{\cal O}(\hbar^{L+1})\\
&=\vf'^a-\phi^a+\sum_{l=1}^{L}\hbar^l\,
\frac{\delta\bUp_l}{\delta\g_a}(\vf',\omega',\ldots)\\
&=-\frac{\delta\varPsi_{L}}{\delta\g_a}(\vf',\omega',\ldots)
=\tilde\vf_{L}^a(\vf',\omega',\ldots)-\phi^a\;,
\end{split}
\ee
where in passing to the fourth line we have assumed that the ${\cal
  O}(\hbar^{L+1})$ terms in (\ref{redefvf}) are adjusted to absorb the (local)
contributions produced by the change of variables in $\bUp_l$, $1\leq
l\leq L-1$. Exactly the same reasoning applies to 
$\delta\varPsi_{L-1}/\delta\z_\a$. 

In the last step, we erase primes on the new variables. Thus, 
we have found the choice of variables 
in the path integral, such that
Eqs.~(\ref{Wgen})---(\ref{reparam1}) are satisfied at the $L$-th loop
order.
This statement extends to all loops by induction. This
completes the proof of the proposition formulated in
Sec.~\ref{sec:2.4} and is the main result of this work. $\blacksquare$

%%%%%%%%%%%%%%%%%%%%%%%%%%%%
\section{Counterterms and nonlinear field renormalization in
$O(N)$ model: Explicit one-loop calculation}
\label{sec:woex}
%%%%%%%%%%%%%%%%%%%%%%%%%%%%

As an illustration of the above formalism we study one-loop
counterterms in the
(1+1)-dimensional $O(N)$-invariant sigma model. 
In particular, we will see the necessity of a
nonlinear
field renormalization to restore the BRST structure.
We start with the action,
\be
\label{sigmaact1}
S=\frac{1}{2 {\gc}^2}\int {\rm d}^2x\; \d_\m n_i \d^\m n^i\;,
\ee
where $i=1,\ldots,N$; ${\gc}$ is the coupling constant 
and the scalar fields $n^i(x)$ are subject to the
constraint,
\be
\label{nsquare}
n^2\equiv \delta_{ij} n^i n^j=1\;.
\ee
The latter can be solved by expressing 
\be
\label{nvf}
n^i=\frac{\vf^i}{\sqrt{\vf^2}}\;,
\ee
where the fields $\vf^i(x)$ are unconstrained. The price to pay is
the appearance of a gauge symmetry corresponding to the pointwise rescaling
of $\vf^i$,
\be
\label{vfchange}
\delta_{\varepsilon}\varphi^{i}(x)=\varphi^{i}(x)\,\varepsilon(x),
\ee
where $\ve(x)$ is an arbitrary function. Clearly, the transformation
(\ref{vfchange}) leaves $n^i(x)$, and hence the action, invariant. In
terms of $\vf^i$ the action reads,  
\begin{align}
\label{sigmaact2}
S[\varphi]=\frac{1}{2{\gc}^2}\int{\rm d}^2x\,\left\{\frac{1}{\varphi^2}\left[\delta_{ij}
-\frac{\varphi_{i}\varphi_{j}}{\varphi^2}\right]\,
\partial_{\mu}\varphi^{i}\partial^{\mu}\varphi^{j}\right\}.
\end{align}
The gauge generator is linear in the fields, 
\be
\label{sigmagen}
R^a_{~b\a}\mapsto \delta^i_j\,\delta(x-x_1)\,\delta(x-x_2)~,~~~~
P^a_{~\a}=0\;,
\ee
so this model belongs
to the class of theories subject to our renormalization
procedure. 
For the sake of convenience we set the coupling constant ${\gc}$ to
one in what follows. 

The local background-covariant gauge condition
$\chi^{\alpha}(\varphi,\phi)$, the gauge fixing matrix
$O^{\alpha\beta}(\phi)$ and its (nonlocal) inverse can be conveniently
chosen in the form 
\bseq
\begin{align}
&\chi^{\alpha}(\varphi,\phi)\mapsto\chi=
\Box\left(\frac{\phi_i(x)}{\phi^2(x)}\,\big(\varphi^{i}(x)-\phi^{i}(x)\big)\right)
=\Box\left(\frac{\phi(x)\cdot\varphi(x)}{\phi^2(x)}\right),\\ 
&O^{\alpha\beta}(\phi)\mapsto O(x,x')
=-\Box\,\delta(x-x'),\quad O^{-1}_{\alpha\beta}(\phi)\mapsto
O^{-1}(x,x')=-\frac{1}{\Box}\,\delta(x-x'),               
\label{6.O}
\end{align}
\eseq
where we have introduced the notation for the $O(N)$-invariant scalar
product, 
\begin{align} 
A\cdot B=\delta_{ij}A^{i}B^i\equiv A_iB^i\;.
\end{align}
The corresponding anticommuting ghost $\omega^\alpha$ and antighost
$\bar\omega_\alpha$, as well as the Lagrange multiplier $b_\alpha$, are
scalars with respect to the $(1+1)$-dimensional Lorentz transformations and
do not carry any $O(N)$ indices,
$\omega^{\alpha}\mapsto\omega(x)$,
$\bar{\omega}_{\alpha}\mapsto\bar{\omega}(x)$, 
$b_{\alpha}\mapsto b(x)$.
The theory is Abelian, $C^\alpha_{\;\;\beta\gamma}=0$, so
that the BRST transform of the 
ghost field $\omega(x)$ vanishes and the source
$\zeta_\alpha$ does not appear in the gauge-fixed action. 
Nevertheless, we have to keep the source $\zeta_\alpha$ in the gauge fermion to fulfil the
requirement (\ref{Psihattree}). 
Therefore, the
tree level reduced gauge fermion equals 
\bseq
\begin{align}
&\hat\varPsi_0=-\hat\gamma_a(\varphi^a-\phi^a)+\z_\a\omega^\a
=\int {\rm d}^2x\,\big(-\hat\gamma_i\,(\varphi^i-\phi^i)+\z\omega\big)\;,\\
&\hat\gamma_i=\gamma_i-\frac{\phi_i}{\phi^2}\,\Box\bar\omega\, .
\end{align}
\eseq
The background field independent choice (\ref{6.O}) of $O$
considerably simplifies the form of the BRST
action (\ref{Sig}) and moreover simplifies the result of integrating
over the Lagrange multiplier $b_\alpha$. The effect of this
integration is the replacement of the 
$b_\alpha$-dependent terms by
the gauge breaking term quadratic in the gauge condition, after which
the BRST action (\ref{Sig}) takes the form (in condensed
notations)\footnote{We disregard the one-loop functional determinant
  $({\rm Det}\, O)^{-1/2}$ originating from this integration, because
  it is a trivial field-independent normalization constant.} 
\begin{align}
\varSigma_0[\varphi,\omega,\bar{\omega},\phi,\g,
\varOmega]={}&S[\varphi]
+\frac{1}{2}\chi^{\alpha}(\varphi,\phi)\,O^{-1}_{\alpha\beta}\,
\chi^{\beta}(\varphi,\phi)
-\bar{\omega}_{\alpha}\,\chi^{\alpha}_a(\phi)\,R^a_{\;\;\beta}(\varphi)\,
\omega^{\beta}\nonumber\\
&+\gamma_{a}\,R^{a}_{\alpha}(\varphi)\,\omega^{\alpha}
+\varOmega^{a}\,\bar{\omega}_{\alpha}\frac{\delta\chi^{\alpha}(\varphi,\phi) }{\delta\phi^{a}}+\vO^a\g_a\;.
\end{align}
Explicitly, the previous action  reads
\be
\begin{split}
\varSigma_0[\varphi,\omega,\bar{\omega},\phi,\g,\varOmega]={}&
\int{\rm d}^2x\,\Bigg\{
\frac{1}{2}\,G_{ij}\,\partial_{\mu}\varphi^{i}\partial^{\mu}\varphi^{j}
-\frac{1}{2}\,\frac{\varphi\cdot \phi}{\phi^2}\,
\Box\left(\frac{\varphi\cdot \phi}{\phi^2}\right)
-\frac{\varphi\cdot\phi}{\phi^2}\,(\Box\,\bar{\omega})\,\omega
\\
&\qquad
+(\gamma\cdot\varphi)\omega
+\left(\frac{\varOmega\cdot\varphi}{\phi^2}
-2\frac{(\varphi\cdot\phi)\,
(\varOmega\cdot\phi)}{(\phi^{2})^2}\right)\,
\Box\, \bar{\omega}
+\vO\cdot\g\Bigg\}.               
\end{split}
\label{actionred}
\ee
Here $G_{ij}$ denotes the metric of the target
manifold,
\begin{align}
G_{ij}={}&\frac{P_{ij}}{\varphi^2},\qquad
P_{ij}=\delta_{ij}-\frac{\varphi_{i}\varphi_{j}}{\varphi^2}\;,
\end{align}
and $P_{ij}$ is a projector along the
directions orthogonal to $\varphi^j$. All terms in the Lagrangian have
mass dimension 2 if the dimensions of the fields are chosen as,
\be
\label{sigmadins}
[\vf]=[\phi]=[\omega]=[\bar\omega]=[\vO]=0~,~~~~[\g]=2\;.
\ee
The theory is renormalizable, hence all divergences also have
dimension 2. This implies that the renormalized gauge fermion
$\hat\varPsi$ should remain linear in $\hat\g_i$ and
independent\footnote{Recall that $\hat\varPsi$ has ghost number
  $(-1)$, whereas the fields have ghost charges (\ref{Qgh}). Note also
that the contribution involving $\z$ in $\hat\varPsi$ does not get
renormalized since $\z$ does not appear in the action.} of
$\vO^i$, as in other renormalizable examples encountered in
Sec.~\ref{sec:examples}. On the other hand, due to the zero mass dimension
of the gauge fields,  we expect that it will have nonlinear
dependence on $\vf^i$ and $\phi^i$. These expectations are
confirmed below by an explicit calculation. 

The one-loop effective action of the model is given by the functional
supertrace, 
\begin{align}
\label{sigG1}
\varGamma_1=\frac12\,{\rm STr}\,\big(\log\,F_{IJ}\big)\;,
\end{align}
where $F_{IJ}$ 
is the inverse propagator of the theory. The latter
is given by the second order mixed (left and right) functional
derivatives of the action with respect to the full set of
boson-fermion fields of the theory
$\varPhi^{I}(x)=(\varphi^{i}(x),\,\,\omega(x),\,\,\bar{\omega}(x))$ 
\begin{align}
F_{IJ}\,\delta(x-x')=
\frac{\overset{\rightarrow}{\delta}}{\delta\varPhi^{I}(x)}
\varSigma_0[\varphi,\omega,\bar\omega,\phi,\gamma,\vO]
\frac{\overset{\leftarrow}{\delta}}{\delta\varPhi^{J}(x')}. \label{FlucOp}
\end{align}
This second order differential operator acting in the space of
perturbations of the fields $\delta\varPhi^J$ has the form, 
\begin{align}
F_{IJ}=D_{IJ}\,\Box+2\,\varGamma_{IJ}^{\mu}\,
\partial_{\mu}+\varPi_{IJ}\;.                           
\label{F}
\end{align}
The expressions for the
matrix valued coefficients $D_{IJ}$, $\varGamma_{IJ}^{\mu}$ and
$\varPi_{IJ}$ are given in Appendix~\ref{app:sigtech}.

The divergent part of (\ref{sigG1}) for a general operator of the form
(\ref{F})
is easily obtained by the heat
kernel method as a local functional of the operator coefficients
\cite{DeWitt:1967ub,DeWitt_book,Barvinsky-Vilkovisky}. 
First, the inverse propagator is converted into 
the form of a {\em covariant}
d'Alembertian, 
\begin{align}
\mbox{\boldmath$F$}^I_{\;J}=-({\cal D}_{\mu}{\cal D}^{\mu})^I_{\;J}
+\mbox{\boldmath$P$}^I_{\;J},               
\label{hatF0}
\end{align}
built in terms of covariant derivatives ${\cal D}_\mu$ with some
generic connection
$\mbox{\boldmath$\varGamma$}_{\mu}=\mbox{\boldmath$\varGamma$}_{\mu\;J}^{~~I}$. 
These 
covariant derivatives  act in the linear space of fields
$\varPhi=\varPhi^I(x)$ and field matrices
$\mbox{\boldmath$X$}=\mbox{\boldmath$X$}^I_{\;J}(x)$ as 
\begin{align}
{\cal D}_{\mu}\varPhi=\partial_{\mu}\varPhi
+\mbox{\boldmath$\varGamma$}_{\mu}\varPhi,\quad
{\cal D}_{\mu}\mbox{\boldmath$X$}=\partial_{\mu}\mbox{\boldmath$X$}
+[\,\mbox{\boldmath$\varGamma$}_{\mu},
\mbox{\boldmath$X$}\,].                                 \label{covder}
\end{align}
In the case of a
$(1+1)$-dimensional flat spacetime the one-loop divergence takes a
particularly simple form: it depends only on
the potential term $\mbox{\boldmath$P$}$ of this operator 
\begin{align}
 \frac12\,{\rm STr}\,\log\mbox{\boldmath$F$} \big |_{\infty}
=\frac{1}{4\pi(2-d)}
\int{\rm d}^2x\,\text{str}\,\mbox{\boldmath$P$}.
\label{1Gamma}
\end{align}
Here $\text{str}$ is the matrix supertrace over indices $I$,
\be
\text{str}\,\mbox{\boldmath$P$}=\sum_I(-1)^{\epsilon_I}
\mbox{\boldmath$P$}^I_{\;I}\;, 
\ee
where $\epsilon_I=0,1$ is the Grassmann parity of the matrix entry
labeled by the index $I$. We used dimensional regularization to capture the divergence in  the limit $d\to 2$. 

We convert (\ref{F}) into the form (\ref{hatF0}) by canonically
normalizing the second-order derivative term of the inverse
propagator, $F_{IJ}=-D_{IK}\mbox{\boldmath$F$}^K_{\;\,J}$. Then
\begin{align}
\mbox{\boldmath$F$}=-\big(\Box\mbox{\boldmath$1$}
+2\mbox{\boldmath$\varGamma$}^\mu\,\partial_\mu
+\mbox{\boldmath$\varPi$}\big)\;,\quad
(\mbox{\boldmath$\varGamma$}^\mu)^I_{\;J}
=D^{IK}\varGamma_{KJ}^{\mu},\quad
\mbox{\boldmath$\varPi$}^{I}_{\;J}=D^{IK}\,\varPi_{KJ},  
\label{hatF}
\end{align}
where $D^{IK}$ is the inverse of the matrix $D_{IJ}$,
$D^{IK}D_{KJ}=\delta^I_J$. Next, the first-order derivative
term of (\ref{hatF}) is absorbed 
into the covariant derivative (\ref{covder}) with
the connection $\mbox{\boldmath$\varGamma$}_\mu$. As a result, the
operator (\ref{hatF}) takes the form (\ref{hatF0}) with
$\mbox{\boldmath$P$}=-\mbox{\boldmath$\varPi$}+ 
\partial_{\mu}\mbox{\boldmath$\varGamma$}^{\mu}
+\mbox{\boldmath$\varGamma$}^{\mu}\mbox{\boldmath$\varGamma$}_{\mu}$,
so that finally the one-loop divergence reads 
\begin{align}
\varGamma_{1,\infty}=-\frac{1}{4\pi(2-d)}\int{\rm
  d}^2x\,\text{str}\,(\mbox{\boldmath$\varPi$} 
-\mbox{\boldmath$\varGamma$}^{\mu}
\mbox{\boldmath$\varGamma$}_{\mu}),                  \label{div}
\end{align}
where we have dropped the total derivative term\footnote{We also disregard the
  ultralocal contribution of the transition from $F_{IJ}$ to
  $\mbox{\boldmath$F$}$, 
$$\text{STr}\log\,F_{IJ}=\text{STr}\log\mbox{\boldmath$F$}+\text{STr}\log
(-D_{IJ})=\text{STr}\log\mbox{\boldmath$F$}+\delta(0)(...)\;,$$ 
which might be canceled by an appropriate local contribution of the measure in the path integral and anyway vanishes in dimensional regularization.} $\partial_\mu({\rm str}\,
\mbox{\boldmath$\varGamma$}^{\mu})$.
The matrices $\mbox{\boldmath$\varGamma$}^\m$ and
$\mbox{\boldmath$\varPi$}$ are evaluated in
Appendix~\ref{app:sigtech}. Substituting the corresponding expressions
into (\ref{div}) we obtain,
\begin{align}
&\hbar\varGamma_{1,\infty}=-\frac{\hbar}{ 2\pi(2-d)}\int{\rm
  d}^2x\,
\Bigg\{\frac{N-2}{2}\, G_{ij}\partial_{\mu}\varphi^i\partial^{\mu}\varphi^{j}
+\frac{(\phi^2)^2}{(\varphi\cdot\phi)^2}
(\varphi\cdot\hat{\gamma})\,\omega\nonumber\\
&+\left(\frac{\delta_{ij}}{\varphi^2}
-2\frac{\phi_{i}\varphi_{j}}{(\varphi\cdot\phi)\varphi^2}
+\frac{\phi_i\phi_j}{(\varphi\cdot\phi)^2}\right)
\partial_{\mu}\varphi^i\partial^{\mu}\varphi^{j}
-\left(\frac{\delta_{ij}}{(\varphi\cdot\phi)}
-\frac{\phi_i\varphi_j}{(\varphi\cdot\phi)^2}\right)
\partial_{\mu}\varphi^i\partial^{\mu}\phi^{j}\nonumber\\
&-\left[\frac{\varphi^2}{(\varphi\cdot\phi)}
\delta^{i}_k-2\,\frac{\phi^2}{(\varphi\cdot\phi)^2}\varphi^{i}\phi_{k}
-\frac{\varphi^2}{(\varphi\cdot\phi)^2}\left(\phi^{i}\varphi_k
+\varphi^{i}\phi_k\right)+\frac{\phi^2(\varphi^2
+\phi^2)}{(\varphi\cdot\phi)^3}\varphi^{i}
\varphi_k\right]\varOmega^{k}\hat{\gamma}_i\Bigg\}.      \label{1LoopDiv}
\end{align}
If we set $\phi^i=\vf^i$, $\vO^i=\g_i=\omega=0$, only the first term
in this expression
will survive corresponding to the 
well-known expression for the 1-loop divergence in the $O(N)$-model
(see e.g.~\cite{Polyakov}).  

Let us look at the terms in the last line of (\ref{1LoopDiv}) 
bilinear in $\varOmega^k$
and $\hat\gamma_i$. 
According to Eq.~(\ref{Gdiv0}), they originate from the action of
the operator $\varOmega \delta/\delta\phi$ on the one-loop ($L=1$) quantum
dressing $\mbox{\boldmath $\varUpsilon$}_1$ of the gauge fermion in
(\ref{PsiL}). Clearly, we are in the situation when this dressing 
is independent of $\vO$ and linear in $\hat\g$, 
\[
\mbox{\boldmath
  $\varUpsilon$}_1=\hat\gamma_a\mbox{\boldmath
  $u$}_1^a(\varphi,\phi)\;.
\] 
Therefore, the terms bilinear in $\varOmega^k$ and $\hat\gamma_i$
should be identified with
$\varOmega^a\hat\gamma_b\,\delta\mbox{\boldmath $u$}_1^b/\delta\phi^a$,
or 
\begin{align}
\frac{\partial\mbox{\boldmath $u$}_1^i(\varphi,\phi)}{\partial\phi^k}\!\!=\!
\frac1{2\pi(2-d)}\!\!\left[\frac{\varphi^2}{\varphi\cdot\phi}
\delta^{i}_k-\frac{2\phi^2}{(\varphi\cdot\phi)^2}\varphi^{i}\phi_{k}
-\frac{\varphi^2}{(\varphi\cdot\phi)^2}\left(\phi^{i}\varphi_k
+\varphi^{i}\phi_k\right)+\frac{\phi^2(\varphi^2
+\phi^2)}{(\varphi\cdot\phi)^3}\varphi^{i}
\varphi_k\right]\!.
\end{align}
One can check that a nontrivial integrability condition for this
equation is satisfied, and the solution reads 
\begin{align}
\mbox{\boldmath $u$}_1^i(\varphi,\phi)=
-\frac1{4\pi(2-d)}\left[\frac{\phi^2(\varphi^2
+\phi^2)}{(\varphi\cdot\phi)^2}\,\varphi^{i}
-\frac{2\varphi^2}{(\varphi\cdot\phi)}\phi^{i}\right].
\end{align}
According to (\ref{Psirenorm}),
(\ref{UL})
 this function generates
 the one-loop field
renormalization, 
\be
\label{sigrepar}
\varphi^i\mapsto \tilde\vf^i_1=
\varphi^i+\hbar\mbox{\boldmath$u$}_1^i(\varphi,\phi)\;.
\ee
Notice that this renormalization is essentially
nonlinear. Still, it is covariant with respect to simultaneous
gauge transformations of both quantum and background fields, as it
should~be. 

It remains to be shown that the rest of the terms in (\ref{1LoopDiv})
recover the correct BRST structure of the renormalized action after
the field redefinition (\ref{sigrepar}). We
observe
that the first term of (\ref{1LoopDiv})
is the gauge invariant counterterm -- proportional to the classical
action, 
\begin{align}
\mbox{\boldmath $S$}_1=-\frac{N-2}{ 4\pi(2-d)}\int{\rm d}^2x\,
G_{ij}\partial_{\mu}\varphi^i\partial^{\mu}\varphi^{j}\;.
\end{align}
The second term bilinear in $\hat\gamma_i$ and $\omega$ can be
represented as the sum of two terms: 
\bseq
\begin{align}
&\hat\gamma_a\frac{\delta R^a_{~\alpha}}{\delta\varphi^b}\,
\mbox{\boldmath $u$}_1^b\omega^\alpha\!=\!
\int\!\! {\rm d}^2x\, (\hat\g\cdot\mbox{\boldmath $u$}_1)\omega
=
\frac1{4\pi(2-d)}\int\!{\rm d}^2x\,\bigg[\!\!
-\!\frac{\phi^2(\vf^2+\phi^2)}{(\vf\cdot\phi)^2}
(\hat\g\cdot\vf)
+\frac{2\vf^2}{(\vf\cdot\phi)}(\hat\g\cdot\phi)\bigg]\,\omega
\,,\\
&-\hat\gamma_a\frac{\delta\mbox{\boldmath $u$}_1^a}{\delta\varphi^b}
R^b_{~\alpha}\omega^\alpha\!
=\!-\!\int\!\! {\rm d}^2x\,\hat\g_i\frac{\d\mbox{\boldmath
    $u$}_1^i}{\d\vf^k}\vf^k\omega
=
\frac{1}{4\pi(2\!-\!d)}\!\int \! {\rm d}^2x\bigg[
\frac{\phi^2(\vf^2\!-\!\phi^2)}{(\vf\cdot\phi)^2}
(\hat\g\cdot\vf)
\!-\!\frac{2\vf^2}{(\vf\cdot\phi)}(\hat\g\cdot\phi)\bigg]
\,\omega.
\end{align}
\eseq
Finally, the second line of (\ref{1LoopDiv}) coincides with the change
of the classical action under the field reparametrization
(\ref{sigrepar}), 
\be
\begin{split}
\frac{\delta S}{\delta\varphi^a}\,\mbox{\boldmath $u$}_1^a=
-\frac1{ 2\pi(2-d)}\int{\rm d}^2x\,\bigg[
&\left(\frac{\delta_{ij}}{\varphi^2}
-2\frac{\phi_{i}\varphi_{j}}{(\varphi\cdot\phi)\varphi^2}
+\frac{\phi_i\phi_j}{(\varphi\cdot\phi)^2}\right)
\partial_{\mu}\varphi^i\partial^{\mu}\varphi^{j}\\
&-\left(\frac{\delta_{ij}}{(\varphi\cdot\phi)}
-\frac{\phi_i\varphi_j}{(\varphi\cdot\phi)^2}\right)
\partial_{\mu}\varphi^i\partial^{\mu}\phi^{j}\bigg].
\end{split}
\ee
With the field reparametrization \eqref{sigrepar} we therefore have
\begin{align}
&\varSigma_0\,\big|_{\,\varphi\to\varphi+\hbar{\mbox{\footnotesize\boldmath $u$}}_1}=
S+\mbox{\boldmath $Q$}\,\varPsi_0
+\hbar\left(\frac{\delta S}{\delta\varphi^a}\mbox{\boldmath $u$}_1^a
+b_\alpha\chi^\alpha_a\mbox{\boldmath $u$}_1^a
  +\hat\gamma_a\frac{\delta
    R^a_{~\alpha}}{\delta\varphi^b}\,\mbox{\boldmath $u$}_1^b\omega^\alpha 
+\varOmega^a\frac{\delta\chi^\alpha_b}{\delta\phi^a}\mbox{\boldmath
  $u$}_1^b\bar\omega_\alpha\right)+{\cal O}(\hbar^2), 
\\
&\hbar\varGamma_{1,\infty}\big|_{\,\varphi\to\varphi
+\hbar{\mbox{\footnotesize\boldmath $u$}}_1}=
\hbar\left(\mbox{\boldmath $S$}_1
+\frac{\delta S}{\delta\varphi^a} \mbox{\boldmath $u$}_1^a
+\varOmega^a\hat\gamma_b\frac{\delta\mbox{\boldmath $u$}_1^b}{\delta\phi^a}
+\hat\gamma_a\frac{\delta
  R^a_{~\alpha}}{\delta\varphi^b}\,\mbox{\boldmath
  $u$}_1^b\omega^\alpha
-\hat\gamma_a \frac{\delta\mbox{\boldmath
    $u$}_1^a}{\delta\varphi^b}R^b_{~\alpha}\omega^\alpha\right)+{\cal O}(\hbar^2).
\end{align}
Thus, the renormalized action reads
\begin{align}
\varSigma_1\equiv\big[\,\varSigma_0
-\hbar\varGamma_{1,\infty}\big]_{\,\varphi\to\varphi
+\hbar{\mbox{\footnotesize\boldmath $u$}}_1}=S[\vf]-\hbar\,
\mbox{\boldmath $S$}_1[\vf]
+\mbox{\boldmath $Q$}\,\big(\varPsi_0-\hbar\bUp_1\big)
+{\cal O}(\hbar^2)\;,
\end{align}
where in the expression for $\mbox{\boldmath $Q$}\,\bUp_1$ we took
into account the dependence of
$\hat\gamma_a=\gamma_a-\bar\omega_\alpha\chi^\alpha_a(\phi)$ on 
$\bar\omega$ and
$\phi$. This BRST structure of the one-loop renormalization is in
full agreement with (\ref{BRSTSL}) --- the renormalized gauge
invariant action $S_1[\vf]=S[\varphi]-\hbar\,\mbox{\boldmath
  $S$}_1[\varphi]$ plus the BRST exact term with the gauge fermion
dressed by a local quantum correction inducing the field
reparameterization.

%%%%%%%%%%%%%%%%%%%%%%%%%%%%
\section{Conclusions and discussion}
\label{sec:conclusions}
%%%%%%%%%%%%%%%%%%%%%%%%%%%%

In this paper we have demonstrated the local BRST structure of
renormalization in a wide class of gauge field theories admitting
background-covariant gauges. Simply stated, we have shown that, for theories
of this class, the renormalization procedure does not spoil gauge
invariance. This class encompasses all standard Einstein--YM--Maxwell
theories, whether renormalizable or not. In this way we reproduce the
classical results concerning renormalization of Einstein--YM theories
and strengthen them for the case of theories with Abelian
subgroups. Other representatives of the class covered by our analysis
are non-relativistic YM--Maxwell theories and projectable Ho\v rava
gravity. This offers the first demonstration of the BRST structure of
projectable Ho\v rava gravity which completes the proof of its
renormalizability. 
The previous list of applications of our results is
certainly not exhaustive. As suggested by the example considered in
Sec.~\ref{sec:woex}, they can be useful for studying various
$\sigma$-models and other theories where gauge invariance is
introduced as a tool to resolve the complicated structure of the field
configuration space.   

Our argument makes essential use of the
background fields $\phi$. With a suitable choice of the gauge condition they
allowed us to introduce an additional gauge invariance with respect
to {\em background} gauge transformations.
We then extended the BRST construction with an auxiliary anticommuting
source $\varOmega$ controlling the dependence of the gauge-fixing term
on the background fields. The counterterms generated by
renormalization were shown to belong to the local 
cohomology of the extended
BRST operators on the space of functionals polynomial in $\varOmega$
and the Faddeev--Popov ghosts. Our key observation is that the
presence of linearly realized background-gauge invariance allows one to
split the computation of this cohomology into several steps involving
cohomologies of a few simpler operators. By completing these steps we
have concluded that the counterterms split into a BRST exact
piece and a 
local gauge invariant functional $\mbox{\boldmath $S$}[\,\varphi\,]$ 
depending only on
the  dynamical -- ``quantum''-- fields which 
renormalizes the physical action of the system.
Our derivation is self-contained and does not rely on any power
counting considerations. We have discussed the simplifications that
appear if such considerations apply.
Our results agree with those of \cite{Anselmi:2013kba} whenever
they overlap.

We have discussed in
detail the local field redefinition bringing the renormalized action into
the BRST form. This field redefinition,
which in simple models has a multiplicative linear nature, becomes essentially nonlinear in generic
theories, as we illustrated with an
explicit example (Sec.~\ref{sec:woex}). 
Despite this complication, it preserves a universal structure:
At any order in the loop expansion, the renormalized quantum
fields are generated by Eq.~(\ref{reparam1}) with the {\em local} generating
functional $\varPsi$. The latter is 
identical to the gauge fermion appearing  in the
exact part $\mbox{\boldmath$Q$}\,\varPsi$ of the full BRST action 
$\varSigma=S[\,\varphi\,]+\mbox{\boldmath$Q$}\varPsi$  dressed by loop
corrections. 
This property provides a systematic algorithm to construct the
field redefinition order by order in perturbation theory. What one
needs to do is just to determine $\varPsi$ from the part of the
counterterm containing the BRST sources and background fields. This
procedure becomes particularly efficient when there are additional
constraints, e.g. due to power counting, that prevent $\varPsi$ from
depending on the BRST source $\varOmega$ associated to background
fields. In that case, our results imply that the $\varOmega$-dependent
part of the counterterm has the form $\vO\,\delta\varPsi/\delta\phi$
(see the definition of $\bQ$ in (\ref{Qlong})). Therefore, $\varPsi$ can be
found by simply integrating the coefficient in front of $\vO$ with
respect to the background fields.
In terms of renormalized fields, the 
 physical part $S[\,\varphi\,]$ of the renormalized action becomes gauge
 invariant. Thus, the divergences contained in $S[\,\varphi\,]$ have
 the same structure as the terms in the tree-level action and are
 absorbed by renormalization of the physical coupling constants.

It is worth reviewing the  various assumptions about the
gauge algebra that entered  into our derivation. An essential assumption
is the linearity
of the gauge generators in the gauge fields which allows one to easily construct
background-covariant gauge conditions.
Moreover, 
the linearity of the resulting background-gauge covariance
is crucial for its preservation at the quantum level. Another
essential requirement is local completeness of the gauge generators
expressed by Eqs.~(\ref{Sannih}), (\ref{Xrepr}). This plays an
important role in the homological analysis of 
the
Koszul--Tate differential performed in 
\cite{Henneaux:1990rx,Vandoren:1993bw} and whose results we used 
in Sec.~\ref{sec:5.2}. On the other hand, it appears
likely that the irreducibility condition \ref{(ii)} from Sec.~\ref{sec:2.1}
can be relaxed at the price of considerably complicating the
ghost sector. Indeed, the main steps in the proof in Sec.~\ref{sec:5}
would be unchanged, including the results of
\cite{Henneaux:1990rx,Vandoren:1993bw} that are straightforwardly
generalized to the reducible case.
Finally, we assumed the gauge algebra to close off-shell which allowed
us to use the standard BRST construction for the gauge fixing. It
would be interesting to extend our analysis to gauge theories with
open algebras. The close connection between our
approach and the Batalin--Vilkovisky 
generalization of the BRST
formalism to open algebras \cite{Batalin:1981jr,BV3} makes the existence
of such extension quite 
plausible.  

Though we have not addressed this topic in the present paper, we
believe that our method can be efficiently applied to renormalization
of composite operators in gauge theories. Another aspect of 
renormalization that has been left outside the scope
of this paper is that of quantum gauge anomalies. These are known to be
related to BRST cohomologies with non-vanishing ghost number. It would
be interesting to see if the background-field approach along the lines
developed here can shed new light on this topic. We
leave this study for future.

\paragraph*{Acknowledgments} We are indebted to Frank Ferrari,
Elias Kiritsis and Igor Tyutin for stimulating discussions. We thank
Ioseph Buchbinder and Marc Henneaux for valuable comments on the first
version of the paper. 
A.B. and
C.S. are grateful
for hospitality of the CERN Theoretical Physics Department where part of this work
has been completed.  
This work was 
supported by the RFBR grant No.17-02-00651 (A.B. and S.S.), the Tomsk State
University Competitiveness Improvement Program (A.B.),  
the Tomalla Foundation (M.H.-V.) and the Swiss
National Science Foundation (S.S.).

\appendix
\renewcommand{\theequation}{\Alph{section}.\arabic{equation}}

%%%%%%%%%%%%%%%%%%%%%%%%%%%%
\section{Derivation of Slavnov-Taylor and Ward identities}
\label{app:STW}
%%%%%%%%%%%%%%%%%%%%%%%%%%%%

To obtain the Slavnov-Taylor identity (\ref{STW}), note that the
total action including the source term in the exponential of
(\ref{Wgen}) can be written in a BRST invariant form. For this
purpose we introduce the ``doubly extended'' BRST operator 
\begin{align}
\mbox{\boldmath$Q$}_{\rm ext}=\mbox{\boldmath$s$}
+\varOmega\frac\delta{\delta\phi}
-J\frac\delta{\delta\gamma}+\bar\xi\frac\delta{\delta\zeta}
+\xi\frac\delta{\delta y}
,\qquad \mbox{\boldmath$Q$}_{\rm ext}^2=0\;,
\end{align}
and notice that the source term in the non-minimal sector can also be
rewritten as a BRST-exact expression, 
\begin{align}
\xi\bar\omega+yb=\left(\mbox{\boldmath$s$}
+\xi\frac\delta{\delta y}\right)\,y\bar\omega
=\mbox{\boldmath$Q$}_{\rm ext}\,(y\bar\omega)\;,
\end{align}
where for brevity we omit the condensed indices of all
quantities. Therefore, the total BRST action including all sources
takes a compact form
in terms of the {\em extended} gauge fermion~$\varPsi_{\rm ext}$, 
\begin{align}
\varSigma_{\rm ext}=\varSigma-J\frac{\delta\varPsi}{\delta\gamma}
+\bar\xi\frac{\delta\varPsi}{\delta\zeta}+\xi\bar\omega+y b=S+\mbox{\boldmath$Q$}_{\rm ext}\varPsi_{\rm ext},
\quad \varPsi_{\rm ext}\equiv \varPsi+y\bar\omega,
\end{align}
and the path integral for the generating functional (\ref{Wgen}) reads
\begin{align}
e^{-W/\hbar}=\int d\varPhi\,e^{-\varSigma_{\rm ext}/\hbar}.  \label{A4}
\end{align}
Clearly,
\begin{align}
\mbox{\boldmath$Q$}_{\rm ext}\,e^{-\varSigma_{\rm ext}/\hbar}=0,
\end{align}
or
\begin{align}
\left(-J\frac\delta{\delta\gamma}+\bar\xi\frac\delta{\delta\zeta}
+\xi\frac\delta{\delta y}
+\varOmega\frac\delta{\delta\phi}\right)\,e^{-\varSigma_{\rm ext}/\hbar}=
-\mbox{\boldmath$s$}\,e^{-\varSigma_{\rm ext}/\hbar},
\end{align}
whence
\begin{align}
\left(-J\frac\delta{\delta\gamma}+\bar\xi\frac\delta{\delta\zeta}
+\xi\frac\delta{\delta y}
+\varOmega\frac\delta{\delta\phi}\right)\,e^{-W/\hbar}=
-\int d\varPhi\,\mbox{\boldmath$s$}\,e^{-\varSigma_{\rm ext}/\hbar}.
\end{align}
The path integral here has the form,
\begin{equation}
\label{B6}
\int d\varPhi\,\big(\mbox{\boldmath$s$} \varPhi^I\big)\,\frac\delta{\delta\varPhi^I}\,
e^{-F[\Phi]}=
-\int  d\varPhi\,\Big(\frac{\delta}{\delta\varPhi^I}\,
\mbox{\boldmath$s$}\varPhi^I(\varPhi)\Big)\,
e^{-F[\Phi]}\;,
\end{equation}
where the expression in brackets on the r.h.s. is the variation of the
integration measure $ d\varPhi$ under the BRST variation of the
fields. It vanishes according to the assumption of anomaly-free
regularization and we arrive at Eq.~(\ref{STW}).

For the derivation of the Ward identity (\ref{WardW}), we introduce,
together with the quantum fields $\varPhi$ and background fields $\phi$, also
the collective notations for all the sources 
\begin{align}
{\cal J}=J_a,\bar\xi_\alpha,\xi^\alpha,y^\alpha,\gamma_a,\zeta_\alpha,\varOmega^a.
\end{align}
Then, in view of our choice of background-covariant gauge conditions, 
$\varSigma_{\rm ext}[\,\varPhi,\phi,{\cal J}\,]$
is invariant with respect to the background-gauge 
transformations\footnote{Note that the
  invariance of the source term 
  $-J_a\delta\varPsi/\delta\gamma_a=J_a(\tilde\varphi_L-\phi)^a$ relies
  on the homogeneity of the linear transformation law for
  $(\tilde\varphi_L-\phi)^a$ contragredient to the transformation of
  $J_a$ in (\ref{btrans1}).} 
(\ref{btrans}), (\ref{btrans1}) supplemented by
\be
\label{BGTsources}
\delta_\varepsilon
J_a=-J_b R^b_{~a\alpha }\varepsilon^\alpha~,~~~
\delta_\ve \bar\xi_\a=\bar\xi_\b C^\b_{~\a\g}\ve^\g~,~~~
\delta_\ve \xi^\a=-C^\a_{~\b\g}\xi^\b\ve^\g~,~~~
\delta_\ve y^\a=-C^\a_{~\b\g}y^\b\ve^\g\;.
\ee
We have,
\begin{align}
\delta_\varepsilon\varSigma_{\rm ext} =\left(\delta_\varepsilon\varPhi\frac\delta{\delta\varPhi}+
\delta_\varepsilon\phi\frac\delta{\delta\phi}
+\delta_\varepsilon{\cal J}
\frac\delta{\delta{\cal J}}\right)
\varSigma_{\rm ext}=0.                      \label{A9}
\end{align}
Next, we perform  the change of integration
variables $\varPhi\to\varPhi+\delta_\varepsilon\varPhi$  in the path integral (\ref{A4}). If, as we did before,  we
disregard the gauge variation of the integration measure, we obtain
the following
integral identity,
\begin{align}
\int d\varPhi\,\delta_\varepsilon\varPhi\,\frac{\delta\varSigma_{\rm ext}}{\delta\varPhi}\,e^{-\varSigma_{\rm ext}/\hbar}=0.
\end{align}
On account of Eq. ({\ref{A9}), its l.h.s. equals
\begin{align}
\int d\varPhi\,\left(\delta_\varepsilon\phi\frac\delta{\delta\phi}
+\delta_\varepsilon{\cal J}\frac\delta{\delta{\cal J}}\right)\,e^{-\varSigma_{\rm ext}/\hbar}=
\left(\delta_\varepsilon\phi\frac\delta{\delta\phi}
+\delta_\varepsilon{\cal J}\frac\delta{\delta{\cal J}}\right)\,e^{-W/\hbar},
\end{align}
because the operator $\delta_\varepsilon\phi\,\delta/\delta\phi
+\delta_\varepsilon{\cal J}\delta/\delta{\cal J}$ is independent of
the integration fields $\varPhi$ and can be commuted with the
integration sign. Therefore, 
\begin{align}
\left(\delta_\varepsilon\phi\frac\delta{\delta\phi}
+\delta_\varepsilon{\cal J}\frac\delta{\delta{\cal J}}\right)\,W=0,
\end{align}
which in view of the expressions (\ref{btrans}), (\ref{btrans1}) and
(\ref{BGTsources}) 
for $\delta_\varepsilon\phi$ and $\delta_\varepsilon{\cal J}$ is just
the expression~(\ref{WardW}).

%%%%%%%%%%%%%%%%%%%%%%%%%%%%
\section{Homology of the operator $\vO\delta/\delta\phi$}
\label{app:coh}
%%%%%%%%%%%%%%%%%%%%%%%%%%%%

In this Appendix we prove the statement used in Sec.~\ref{sec:5.1} that  the cohomology of the
operator $\vO\delta/\delta\phi$ on the space of 
local functionals vanishing at $\vO=0$ is trivial.

{\it Lemma:} Let $X[\vf,\phi,\vO,\ldots]$ be a local
functional of the gauge fields $\vf^a$, background fields
$\phi^a$, anticommuting BRST sources $\vO^a$ and, possibly, other
fields represented by dots. Assume that $X$ vanishes at $\vO^a=0$,
\be
\label{Xvan}
X\big|_{\vO=0}=0\;,
\ee  
that it is invariant under background-gauge
transformations and satisfies the equation
\be
\label{Xeq}
\vO^a\frac{\delta X}{\delta\phi^a}=0\;.
\ee 
Then there exists a local functional $Y$, invariant under background-gauge transformations, such that 
\be
\label{Xsol}
X=\vO^a\frac{\delta Y}{\delta\phi^a}\;.
\ee

\emph{Proof:} The functional $Y$ is constructed explicitly as,
\be
\label{Yexprs}
Y=(\phi^a-\vf^a)\frac{\delta}{\delta\vO^a}
\int_0^1 \frac{{\rm d}z}{z}X[\vf,\vf+z(\phi-\vf),z\vO,\ldots]\;,
\ee
where the arguments of $X$ represented by dots are left
untouched. According to the assumption (\ref{Xvan}), this expression
indeed provides a regular functional. Notice that if $X$ is local, so
is (\ref{Yexprs}). Moreover, $Y$ inherits background gauge invariance 
from $X$
due to the linearity of background-gauge
transformations. It remains to
demonstrate (\ref{Xsol}). Using the anticommutator
\be
\label{anticom}
\vO^a\frac{\delta}{\delta\phi^a}~
(\phi^b-\vf^b)\frac{\delta}{\delta\vO^b}
+(\phi^b-\vf^b)\frac{\delta}{\delta\vO^b}
~\vO^a\frac{\delta}{\delta\phi^a}
=\vO^a\frac{\delta}{\delta\vO^a}
+(\phi^a-\vf^a)\frac{\delta}{\delta\phi^a}
\ee
we find
\be
\label{Ydif}
\vO^a\frac{\delta}{\delta\phi^a} Y=
\int_0^1\frac{{\rm d}z}{z}\left(
z\frac{{\rm d~}}{{\rm d}z}X[\vf,\vf+z(\phi-\vf),z\vO,\ldots]\right)
=X[\vf,\phi,\vO,\ldots]\;,
\ee
where we again used (\ref{Xvan}). $\blacksquare$ 

%%%%%%%%%%%%%%%%%%%%%%%%%%%%
\section{Quadratic form for perturbations in the $O(N)$ model}
\label{app:sigtech}
%%%%%%%%%%%%%%%%%%%%%%%%%%%%

Here we summarize the expressions for the coefficients of the operator
(\ref{F}) appearing in the quadratic action for the perturbations 
$\delta\varPhi^I=(\delta\vf^i,\delta\omega,\delta\bar\omega)$ of the
$O(N)$ model. We write these coefficients as
matrices with
 $3\times 3$ block structure.  The coefficient of the
d'Alembertian is the (super)symmetric matrix 
\begin{align}
D_{IJ}={}&
\left(
\begin{array}{ccc}
A_{ij} &
0 &
B_i\\
0&
0&
C\\
-B_{j} &
-C &
0
\end{array}
\right)  \,,                 \label{D}
\end{align}
with the following boson-boson $A_{ij}$, boson-fermion $B_i$ and fermion-fermion $C$ entries
\begin{align}
A_{ij}={}&-G_{ij}-\frac{\phi_i\phi_j}{(\phi^2)^2}=
-\frac{1}{\varphi^2}\left[\delta_{ij}
-\frac{\varphi_i\varphi_j}{\varphi^2}
+\frac{\varphi^2}{\phi^2}\frac{\phi_i\phi_j}{\phi^2}\right], \label{Aij}\\
B_{i}={}&\frac{\phi_i}{\phi^2}\,\omega
+\frac{\varOmega_{i}}{\phi^2}-2\,
\frac{(\varOmega\cdot\phi)}{(\phi^2)^2}\phi_{i},\qquad~~
C=\frac{\varphi\cdot\phi }{\phi^2}.      \label{BC}
\end{align}
The other two matrix coefficients have the form,
\begin{align}
&\varGamma_{IJ}^{\mu}=\left(\begin{array}{ccc}
\varGamma^{\mu}_{ij} &
0 &
0\\
0&
0&
0\\
-\partial^{\mu}B_j &
\!\!-\partial^{\mu}C &
0
\end{array}\right),~~~~~
\varPi_{IJ}=\left(\begin{array}{ccc}
\varPi_{ij} &
\hat\gamma_{i} &
0\\
-\hat\gamma_{j}&
0&
0\\
-\Box B_j &
-\Box C &
0
\end{array}\right),
\\
&\varGamma_{ij}^{\mu}=-\frac{1}{2}(G_{ji,k}+G_{ki,j}
-G_{jk,i})\,\partial^{\mu}\varphi^{k}
-\frac{\phi_i}{\phi^2}\,\partial^{\mu}
\left(\frac{\phi_j}{\phi^2}\right),\\
&\varPi_{ij}=-\left(G_{ik,jl}-\frac{1}{2}G_{kl,ij}\right)
\partial_{\mu}\varphi^k\partial^{\mu}\varphi^l-G_{ki,j}
\Box\varphi^k-\frac{\phi_i}{\phi^2}
\Box\left(\frac{\phi_j}{\phi^2}\right),  \label{Wblock}
\end{align}
where
\begin{align}
G_{ij,k}\equiv&\frac{\partial G_{ij}}{\partial\varphi^k}
=-\frac{2\delta_{ij}\vf_k+\delta_{ik}\vf_j+\delta_{jk}\vf_i}{(\vf^2)^2}
+\frac{4\vf_i\vf_j\vf_k}{(\vf^2)^3}
\;,\notag\\
G_{ij,kl}\equiv&\frac{\partial^2
  G_{ij}}{\partial\varphi^k\partial\varphi^l}
=-\frac{2\delta_{ij}\delta_{kl}+\delta_{ik}\delta_{jl}+\delta_{jk}\delta_{il}}
{(\vf^2)^2}-\frac{24\vf_i\vf_j\vf_k\vf_l}{(\vf^2)^4}\notag\\
&+\frac{4}{(\vf^2)^3}\Big(2\delta_{ij}\vf_k\vf_l+\delta_{ik}\vf_j\vf_l
+\delta_{il}\vf_j\vf_k+\delta_{jk}\vf_i\vf_l+\delta_{jl}\vf_i\vf_k
+\delta_{kl}\vf_i\vf_j\Big)\notag
\;.
\end{align}

For the computation of the one-loop divergence of
the effective action, we need the matrices  
$\mbox{\boldmath$\varPi$}$ and
$\mbox{\boldmath$\varGamma$}^{\mu}$ defined in (\ref{hatF}).
This, in turn, 
requires the
inverse $D^{IJ}$ of the matrix  (\ref{D}), which reads 
\begin{align}
D^{IJ}={}&
\left(
\begin{array}{ccc}
A^{ij} &
-A^{ik}B_{k}/C &
0\\
\\
-B_{k}A^{kj}/C&
B_{k}A^{kl}B_l/C^2&
-1/C\\
\\
0 &
1/C &
0
\end{array}\right)\label{InverseD},
\end{align}
where $B_j$ and $C$ are given by (\ref{BC}) and 
\be
A^{ij}={}-\varphi^2\left[\delta^{ij}
-\frac{\left(\varphi^{i}\phi^{j}
+\phi^{i}\varphi^{j}\right)}{\varphi\cdot\phi}
+\frac{\phi^2}{\varphi^2}
\frac{(\varphi^2+\phi^2)}{(\varphi\cdot\phi)^2}
\varphi^{i}\varphi^{j}\right]
\ee
is the
inverse of the matrix $A_{ij}$ defined by (\ref{Aij}),
$A_{il}A^{lj}=\delta^j_i$. Using these expressions we obtain, 
\begin{align}
\mbox{\boldmath$\varGamma$}^\mu={}&
\left(
\begin{array}{ccc}
A^{il}\varGamma^{\mu}_{lk} &
0 &
0\\
\\
(\partial^{\mu}B_k-B_lA^{lm}\varGamma^{\mu}_{mk})/C&
\partial^{\mu}C/C&
0\\
\\
0 &
0 &
0
\end{array}\right).
\end{align}
The diagonal blocks of $\mbox{\boldmath$\varGamma$}_\mu^2$ equal
\begin{align}
(\mbox{\boldmath$\varGamma$}_\mu^2)^{\,i}_{\,j}
=A^{il}\varGamma^{\mu}_{lk}A^{km}\varGamma_{\mu mj},\quad
(\mbox{\boldmath$\varGamma$}_\mu^2)^{\,\omega}_{\,\omega}
=\frac1{C^2}\,(\partial_\mu C)^2,\quad
(\mbox{\boldmath$\varGamma$}_\mu^2)^{\,\bar\omega}_{\,\bar\omega}=0.
\end{align}
For the matrix of the potential term $\mbox{\boldmath$\varPi$}$ 
we need only its diagonal
block elements which read 
\begin{align}
\mbox{\boldmath$\varPi$}^i_j=A^{il}\,\varPi_{lj}+\frac1C\,A^{il}\,B_l\hat\gamma_j,\quad \mbox{\boldmath$\varPi$}^{\,\omega}_{\,\omega}=
-\frac1C\,(\,B_k\,A^{kl}\,\hat\gamma_l -\Box C),\quad
\mbox{\boldmath$\varPi$}^{\,\bar\omega}_{\,\bar\omega}=0. 
\end{align}
Substituting the above results and expressions for $A^{ij}$, $B_i$ and
$C$ into the supertrace of Eq.~(\ref{div}),
$\text{str}\,\mbox{\boldmath$\varPi$}=\mbox{\boldmath$\varPi$}^i_i 
-\mbox{\boldmath$\varPi$}^\omega_\omega
-\mbox{\boldmath$\varPi$}^{\bar\omega}_{\bar\omega}$ and similarly for
$\text{tr}\,\mbox{\boldmath$\varGamma$}_\mu^2$, we arrive at
Eq.~(\ref{1LoopDiv}).


\begin{thebibliography}{99}

%\cite{Becchi:1974md}
\bibitem{Becchi:1974md} 
  C.~Becchi, A.~Rouet and R.~Stora,
  %``Renormalization of the Abelian Higgs-Kibble Model,''
  Commun.\ Math.\ Phys.\  {\bf 42}, 127 (1975); 
%  doi:10.1007/BF01614158
  %%CITATION = doi:10.1007/BF01614158;%%
  %934 citations counted in INSPIRE as of 04 Dec 2015
%\cite{Becchi:1975nq}
%\bibitem{Becchi:1975nq} 
%  C.~Becchi, A.~Rouet and R.~Stora,
  %``Renormalization of Gauge Theories,''
  Annals Phys.\  {\bf 98}, 287 (1976).
%  doi:10.1016/0003-4916(76)90156-1
  %%CITATION = doi:10.1016/0003-4916(76)90156-1;%%
  %1315 citations counted in INSPIRE as of 04 Dec 2015

\bibitem{Tyutin1}
I.V.~Tyutin, Lebedev Institute preprint N39 (1975).

\bibitem{Weinberg}
S.~Weinberg, {\it The Quantum Theory of Fields,} vol. 2: {\it Modern
Applications,} Cambridge University Press (1996).

%\cite{Stelle:1976gc}
\bibitem{Stelle:1976gc}
  K.~S.~Stelle,
  %``Renormalization of Higher Derivative Quantum Gravity,''
  Phys.\ Rev.\ D {\bf 16}, 953 (1977).
 % doi:10.1103/PhysRevD.16.953
  %%CITATION = doi:10.1103/PhysRevD.16.953;%%
  %1164 citations counted in INSPIRE as of 04 fevr. 2016

\bibitem{ZinnJustin}
J.~Zinn-Justin, ``Renormalization of gauge theories.'' In: Rollnik
H., Dietz K. (eds) {\it Trends in Elementary Particle Theory. Lecture Notes
in Physics,} vol 37. Springer, Berlin, Heidelberg (1975).

%\cite{Voronov:1982cp}
\bibitem{Voronov:1982cp} 
  B.~L.~Voronov and I.~V.~Tyutin,
%  ``Formulation Of Gauge Theories Of General Form. I,''
  Theor.\ Math.\ Phys.\  {\bf 50}, 218 (1982)
  [Teor.\ Mat.\ Fiz.\  {\bf 50}, 333 (1982)]; 
%  doi:10.1007/BF01016448
  %%CITATION = doi:10.1007/BF01016448;%%
  %64 citations counted in INSPIRE as of 25 Mar 2017
%\cite{Voronov:1982ur}
%\bibitem{Voronov:1982ur} 
%  B.~l.~Voronov and I.~v.~Tyutin,
%  ``Formulation Of Gauge Theories Of General Form. II. Gauge Invariant Renormalizability And Renormalization Structure,''
  Theor.\ Math.\ Phys.\  {\bf 52}, 628 (1982)
  [Teor.\ Mat.\ Fiz.\  {\bf 52}, 14 (1982)].
%  doi:10.1007/BF01027781
  %%CITATION = doi:10.1007/BF01027781;%%
  %38 citations counted in INSPIRE as of 25 Mar 2017

%\cite{Anselmi:1994ry}
\bibitem{Anselmi:1994ry} 
  D.~Anselmi,
%  ``Removal of divergences with the Batalin-Vilkovisky formalism,''
  Class.\ Quant.\ Grav.\  {\bf 11}, 2181 (1994).
%  doi:10.1088/0264-9381/11/9/005
  %%CITATION = doi:10.1088/0264-9381/11/9/005;%%
  %24 citations counted in INSPIRE as of 25 Mar 2017

%\cite{Gomis:1995jp}
\bibitem{Gomis:1995jp} 
  J.~Gomis and S.~Weinberg,
%  ``Are nonrenormalizable gauge theories renormalizable?,''
  Nucl.\ Phys.\ B {\bf 469}, 473 (1996)
%  doi:10.1016/0550-3213(96)00132-0
  [hep-th/9510087].
  %%CITATION = doi:10.1016/0550-3213(96)00132-0;%%
  %131 citations counted in INSPIRE as of 25 Mar 2017

%\cite{Barnich:1994ve}
\bibitem{Barnich:1994ve} 
  G.~Barnich and M.~Henneaux,
% ``Renormalization of gauge invariant operators and anomalies in
%  Yang-Mills theory,'' 
  Phys.\ Rev.\ Lett.\  {\bf 72}, 1588 (1994)
%  doi:10.1103/PhysRevLett.72.1588
  [hep-th/9312206].
  %%CITATION = doi:10.1103/PhysRevLett.72.1588;%%
  %84 citations counted in INSPIRE as of 25 Mar 2017

%\cite{Barnich:1994mt}
\bibitem{Barnich:1994mt} 
  G.~Barnich, F.~Brandt and M.~Henneaux,
%  ``Local BRST cohomology in the antifield formalism. II. Application to Yang-Mills theory,''
  Commun.\ Math.\ Phys.\  {\bf 174}, 93 (1995)
%  doi:10.1007/BF02099465
  [hep-th/9405194];
  %%CITATION = doi:10.1007/BF02099465;%%
  %172 citations counted in INSPIRE as of 25 Mar 2017
%\cite{Barnich:1995ap}
%\bibitem{Barnich:1995ap} 
%  G.~Barnich, F.~Brandt and M.~Henneaux,
%  ``Local BRST cohomology in Einstein Yang-Mills theory,''
  Nucl.\ Phys.\ B {\bf 455}, 357 (1995)
%  doi:10.1016/0550-3213(95)00471-4
  [hep-th/9505173].
  %%CITATION = doi:10.1016/0550-3213(95)00471-4;%%
  %72 citations counted in INSPIRE as of 25 Mar 2017


%\cite{Vafek:2002jf}
\bibitem{Vafek:2002jf} 
  O.~Vafek, Z.~Tesanovic and M.~Franz,
  %``Relativity restored: Dirac anisotropy in QED(3),''
  Phys.\ Rev.\ Lett.\  {\bf 89}, 157003 (2002)
%  doi:10.1103/PhysRevLett.89.157003
  [cond-mat/0203047];
  %%CITATION = doi:10.1103/PhysRevLett.89.157003;%%
  %26 citations counted in INSPIRE as of 08 May 2017
%\cite{Franz:2002qy}
%\bibitem{Franz:2002qy} 
%  M.~Franz, Z.~Tesanovic and O.~Vafek,
  %``QED(3) theory of pairing pseudogap in cuprates. 1. From D wave superconductor to antiferromagnet via 'algebraic' Fermi liquid,''
  Phys.\ Rev.\ B {\bf 66}, 054535 (2002)
%  doi:10.1103/PhysRevB.66.054535
  [cond-mat/0203333].
  %%CITATION = doi:10.1103/PhysRevB.66.054535;%%
  %128 citations counted in INSPIRE as of 08 May 2017

%\cite{Ardonne:2003wa}
\bibitem{Ardonne:2003wa}
  E.~Ardonne, P.~Fendley and E.~Fradkin,
  %``Topological order and conformal quantum critical points,''
  Annals Phys.\  {\bf 310} (2004) 493
%  doi:10.1016/j.aop.2004.01.004
  [cond-mat/0311466].
  %%CITATION = doi:10.1016/j.aop.2004.01.004;%%
  %118 citations counted in INSPIRE as of 27 Apr 2017
  
%\cite{Roy:2015zna}
\bibitem{Roy:2015zna}
  B.~Roy, V.~Juricic and I.~F.~Herbut,
  %``Emergent Lorentz symmetry near fermionic quantum critical points in two and three dimensions,''
  JHEP {\bf 1604} (2016) 018
%  doi:10.1007/JHEP04(2016)018
  [arXiv:1510.07650 [hep-th]].
  %%CITATION = doi:10.1007/JHEP04(2016)018;%%
  %5 citations counted in INSPIRE as of 02 May 2017
 
 %\cite{Kachru:2008yh}
\bibitem{Kachru:2008yh}
  S.~Kachru, X.~Liu and M.~Mulligan,
  %``Gravity duals of Lifshitz-like fixed points,''
  Phys.\ Rev.\ D {\bf 78} (2008) 106005
%  doi:10.1103/PhysRevD.78.106005
  [arXiv:0808.1725 [hep-th]].
  %%CITATION = doi:10.1103/PhysRevD.78.106005;%%
  %677 citations counted in INSPIRE as of 27 Apr 2017
  
 %\cite{Griffin:2011xs}
\bibitem{Griffin:2011xs}
  T.~Griffin, P.~Horava and C.~M.~Melby-Thompson,
  %``Conformal Lifshitz Gravity from Holography,''
  JHEP {\bf 1205} (2012) 010
%  doi:10.1007/JHEP05(2012)010
  [arXiv:1112.5660 [hep-th]].
  %%CITATION = doi:10.1007/JHEP05(2012)010;%%
  %61 citations counted in INSPIRE as of 27 Apr 2017 

%\cite{Anselmi:2008bt}
\bibitem{Anselmi:2008bt} 
  D.~Anselmi,
%  ``Weighted power counting, neutrino masses and Lorentz violating
%  extensions of the Standard Model,'' 
  Phys.\ Rev.\ D {\bf 79}, 025017 (2009)
%  doi:10.1103/PhysRevD.79.025017
  [arXiv:0808.3475 [hep-ph]];
  %%CITATION = doi:10.1103/PhysRevD.79.025017;%%
  %49 citations counted in INSPIRE as of 25 Mar 2017
%\cite{Anselmi:2009vz}
%\bibitem{Anselmi:2009vz} 
%  D.~Anselmi,
%  ``Standard Model Without Elementary Scalars And High Energy Lorentz
%  Violation,'' 
  Eur.\ Phys.\ J.\ C {\bf 65}, 523 (2010)
%  doi:10.1140/epjc/s10052-009-1211-z
  [arXiv:0904.1849 [hep-ph]].
  %%CITATION = doi:10.1140/epjc/s10052-009-1211-z;%%
  %51 citations counted in INSPIRE as of 25 Mar 2017

%\cite{Iengo:2010xg}
\bibitem{Iengo:2010xg} 
  R.~Iengo and M.~Serone,
  %``A Simple UV-Completion of QED in 5D,''
  Phys.\ Rev.\ D {\bf 81}, 125005 (2010)
%  doi:10.1103/PhysRevD.81.125005
  [arXiv:1003.4430 [hep-th]].
  %%CITATION = doi:10.1103/PhysRevD.81.125005;%%
  %27 citations counted in INSPIRE as of 25 Mar 2017

%\cite{Barvinsky:2017mal}
\bibitem{Barvinsky:2017mal}
  A.~O.~Barvinsky, D.~Blas, M.~Herrero-Valea, D.~V.~Nesterov,
  G.~P\'erez-Nadal and C.~F.~Steinwachs,
  %``Heat kernel methods for Lifshitz theories,''
  arXiv:1703.04747 [hep-th].
  %%CITATION = ARXIV:1703.04747;%%

%\cite{Horava:2008ih}
\bibitem{Horava:2008ih} 
  P.~Horava,
%  ``Membranes at Quantum Criticality,''
  JHEP {\bf 0903}, 020 (2009)
%  doi:10.1088/1126-6708/2009/03/020
  [arXiv:0812.4287 [hep-th]]; 
  %%CITATION = doi:10.1088/1126-6708/2009/03/020;%%
  %536 citations counted in INSPIRE as of 11 Mar 2017
%\cite{Horava:2009uw}
%\bibitem{Horava:2009uw} 
%  P.~Horava,
%  ``Quantum Gravity at a Lifshitz Point,''
  Phys.\ Rev.\ D {\bf 79}, 084008 (2009)
%  doi:10.1103/PhysRevD.79.084008
  [arXiv:0901.3775 [hep-th]].
  %%CITATION = doi:10.1103/PhysRevD.79.084008;%%
  %1404 citations counted in INSPIRE as of 11 Mar 2017

%\cite{Barvinsky:2015kil}
\bibitem{Barvinsky:2015kil} 
  A.~O.~Barvinsky, D.~Blas, M.~Herrero-Valea, S.~M.~Sibiryakov and
  C.~F.~Steinwachs, 
%  ``Renormalization of Ho\v rava gravity,''
  Phys.\ Rev.\ D {\bf 93}, no. 6, 064022 (2016)
%  doi:10.1103/PhysRevD.93.064022
  [arXiv:1512.02250 [hep-th]].
  %%CITATION = doi:10.1103/PhysRevD.93.064022;%%
  %28 citations counted in INSPIRE as of 12 Mar 2017

%\cite{DeWitt:1967ub}
\bibitem{DeWitt:1967ub} 
  B.~S.~DeWitt,
%  ``Quantum Theory of Gravity. 2. The Manifestly Covariant Theory,''
  Phys.\ Rev.\  {\bf 162}, 1195 (1967);
%  doi:10.1103/PhysRev.162.1195
  %%CITATION = doi:10.1103/PhysRev.162.1195;%%
  %1256 citations counted in INSPIRE as of 25 Mar 2017
%\cite{DeWitt:1967uc}
%\bibitem{DeWitt:1967uc} 
%  B.~S.~DeWitt,
%  ``Quantum Theory of Gravity. 3. Applications of the Covariant Theory,''
  Phys.\ Rev.\  {\bf 162}, 1239 (1967).
%  doi:10.1103/PhysRev.162.1239
  %%CITATION = doi:10.1103/PhysRev.162.1239;%%
  %739 citations counted in INSPIRE as of 25 Mar 2017

%\cite{Abbott:1981ke}
\bibitem{Abbott:1981ke} 
  L.~F.~Abbott,
%  ``Introduction to the Background Field Method,''
  Acta Phys.\ Polon.\ B {\bf 13}, 33 (1982).
  %%CITATION = APPOA,B13,33;%%
  %297 citations counted in INSPIRE as of 25 Mar 2017

\bibitem{DeWitt_book}
B.~S.~DeWitt, {\it Dynamical Theory of Groups and Fields,} Gordon and
Breach (1965).

\bibitem{Veltman}
M.~J.~G.~Veltman, ``Quantum Theory of Gravitation,'' 
Conf. Proc. C {\bf 7507281}, 265 (1975).

%\cite{Honerkamp:1972fd}
\bibitem{Honerkamp:1972fd} 
  J.~Honerkamp,
%  ``The Question of invariant renormalizability of the massless Yang-Mills theory in a manifest covariant approach,''
  Nucl.\ Phys.\ B {\bf 48}, 269 (1972).
%  doi:10.1016/0550-3213(72)90063-6
  %%CITATION = doi:10.1016/0550-3213(72)90063-6;%%
  %242 citations counted in INSPIRE as of 25 Mar 2017

%\cite{tHooft:1973bhk}
\bibitem{tHooft:1973bhk} 
  G.~'t Hooft,
  %``An algorithm for the poles at dimension four in the dimensional regularization procedure,''
  Nucl.\ Phys.\ B {\bf 62}, 444 (1973).
%  doi:10.1016/0550-3213(73)90263-0
  %%CITATION = doi:10.1016/0550-3213(73)90263-0;%%
  %505 citations counted in INSPIRE as of 02 Apr 2017

%\cite{tHooft:1974toh}
\bibitem{tHooft:1974toh} 
  G.~'t Hooft and M.~J.~G.~Veltman,
%  ``One loop divergencies in the theory of gravitation,''
  Ann.\ Inst.\ H.\ Poincare Phys.\ Theor.\ A {\bf 20}, 69 (1974).
  %894 citations counted in INSPIRE as of 25 Mar 2017

%\cite{KlubergStern:1974xv}
\bibitem{KlubergStern:1974xv} 
  H.~Kluberg-Stern and J.~B.~Zuber,
%  ``Renormalization of Nonabelian Gauge Theories in a Background
%  Field Gauge. 1. Green Functions,'' 
  Phys.\ Rev.\ D {\bf 12}, 482 (1975).
%  doi:10.1103/PhysRevD.12.482
  %%CITATION = doi:10.1103/PhysRevD.12.482;%%
  %267 citations counted in INSPIRE as of 25 Mar 2017

%\cite{Tyutin:1978ec}
\bibitem{Tyutin:1978ec} 
  I.~V.~Tyutin,
% ``Renormalization of the Background Functional in Nonabelian Gauge
%  Theories,'' 
  Teor.\ Mat.\ Fiz.\  {\bf 35}, 29 (1978).
%  doi:10.1007/BF01032426
  %%CITATION = doi:10.1007/BF01032426;%%
  %1 citations counted in INSPIRE as of 25 Mar 2017

%\cite{Grassi:1995wr}
\bibitem{Grassi:1995wr}
  P.~A.~Grassi,
%  ``Stability and renormalization of Yang-Mills theory with background
%  field method: A Regularization independent proof,'' 
  Nucl.\ Phys.\ B {\bf 462}, 524 (1996)
 % doi:10.1016/0550-3213(96)00017-X
  [hep-th/9505101].
  %%CITATION = doi:10.1016/0550-3213(96)00017-X;%%
  %50 citations counted in INSPIRE as of 05 Feb 2016

%\cite{Ferrari:2000yp}
\bibitem{Ferrari:2000yp} 
  R.~Ferrari, M.~Picariello and A.~Quadri,
  %``Algebraic aspects of the background field method,''
  Annals Phys.\  {\bf 294}, 165 (2001)
%  doi:10.1006/aphy.2001.6198
  [hep-th/0012090].
  %%CITATION = doi:10.1006/aphy.2001.6198;%%
  %28 citations counted in INSPIRE as of 23 Jun 2017

%\cite{Binosi:2011ar}
\bibitem{Binosi:2011ar} 
  D.~Binosi and A.~Quadri,
  %``Slavnov-Taylor constraints for non-trivial backgrounds,''
  Phys.\ Rev.\ D {\bf 84}, 065017 (2011)
%  doi:10.1103/PhysRevD.84.065017
  [arXiv:1106.3240 [hep-th]].
  %%CITATION = doi:10.1103/PhysRevD.84.065017;%%
  %16 citations counted in INSPIRE as of 23 Jun 2017

%\cite{Kallosh:1974yh}
\bibitem{Kallosh:1974yh} 
  R.~E.~Kallosh,
%  ``The Renormalization in Nonabelian Gauge Theories,''
  Nucl.\ Phys.\ B {\bf 78}, 293 (1974).
%  doi:10.1016/0550-3213(74)90284-3
  %%CITATION = doi:10.1016/0550-3213(74)90284-3;%%
  %151 citations counted in INSPIRE as of 25 Mar 2017

%\cite{Arefeva:1974jv}
\bibitem{Arefeva:1974jv} 
  I.~Y.~Arefeva, L.~D.~Faddeev and A.~A.~Slavnov,
%  ``Generating Functional for the s Matrix in Gauge Theories,''
  Theor.\ Math.\ Phys.\  {\bf 21}, 1165 (1975)
  [Teor.\ Mat.\ Fiz.\  {\bf 21}, 311 (1974)].
%  doi:10.1007/BF01038094
  %%CITATION = doi:10.1007/BF01038094;%%
  %70 citations counted in INSPIRE as of 25 Mar 2017

%\cite{Abbott:1980hw}
\bibitem{Abbott:1980hw} 
  L.~F.~Abbott,
%  ``The Background Field Method Beyond One Loop,''
  Nucl.\ Phys.\ B {\bf 185}, 189 (1981).
%  doi:10.1016/0550-3213(81)90371-0
  %%CITATION = doi:10.1016/0550-3213(81)90371-0;%%
  %1044 citations counted in INSPIRE as of 25 Mar 2017

%\cite{Ichinose:1981uw}
\bibitem{Ichinose:1981uw} 
  S.~Ichinose and M.~Omote,
  %``Renormalization Using the Background Field Method,''
  Nucl.\ Phys.\ B {\bf 203}, 221 (1982).
%  doi:10.1016/0550-3213(82)90029-3
  %%CITATION = doi:10.1016/0550-3213(82)90029-3;%%
  %87 citations counted in INSPIRE as of 02 Apr 2017

\bibitem{Barvinsky-Vilkovisky}
A.~O.~Barvinsky and G.~A.~Vilkovisky, ``The effective action in quantum
field theory: two-loop approximation.'' 
In I.~Batalin, C.~J.~Isham and G.~A.~Vilkovisky (eds)  
{\em Quantum Field Theory and
  Quantum Statistics,} {\bf v.1}, 245, Hilger, Bristol (1987). 

%\cite{Anselmi:2013kba}
\bibitem{Anselmi:2013kba} 
  D.~Anselmi,
%  ``Background field method, Batalin-Vilkovisky formalism and
%  parametric completeness of renormalization,'' 
  Phys.\ Rev.\ D {\bf 89}, no. 4, 045004 (2014)
%  doi:10.1103/PhysRevD.89.045004
  [arXiv:1311.2704 [hep-th]].
  %%CITATION = doi:10.1103/PhysRevD.89.045004;%%
  %14 citations counted in INSPIRE as of 25 Mar 2017

%\cite{Binosi:2012pd}
\bibitem{Binosi:2012pd} 
  D.~Binosi and A.~Quadri,
  %``Canonical Transformations and Renormalization Group Invariance in the presence of Non-trivial Backgrounds,''
  Phys.\ Rev.\ D {\bf 85}, 085020 (2012)
%  doi:10.1103/PhysRevD.85.085020
  [arXiv:1201.1807 [hep-th]];
  %%CITATION = doi:10.1103/PhysRevD.85.085020;%%
  %12 citations counted in INSPIRE as of 23 Jun 2017
%\cite{Binosi:2012st}
%\bibitem{Binosi:2012st} 
%  D.~Binosi and A.~Quadri,
  %``The Background Field Method as a Canonical Transformation,''
  Phys.\ Rev.\ D {\bf 85}, 121702 (2012)
%  doi:10.1103/PhysRevD.85.121702
  [arXiv:1203.6637 [hep-th]].
  %%CITATION = doi:10.1103/PhysRevD.85.121702;%%
  %20 citations counted in INSPIRE as of 23 Jun 2017

%\cite{Batalin:1981jr}
\bibitem{Batalin:1981jr} 
  I.~A.~Batalin and G.~A.~Vilkovisky,
%  ``Gauge Algebra and Quantization,''
  Phys.\ Lett.\  {\bf 102B}, 27 (1981);
%  doi:10.1016/0370-2693(81)90205-7
  %%CITATION = doi:10.1016/0370-2693(81)90205-7;%%
  %965 citations counted in INSPIRE as of 25 Mar 2017
%\bibitem{BV2}
%\cite{Batalin:1983wj}
%\bibitem{Batalin:1983wj} 
%  I.~A.~Batalin and G.~a.~Vilkovisky,
%  ``Feynman Rules For Reducible Gauge Theories,''
  Phys.\ Lett.\  {\bf 120B}, 166 (1983);
%  doi:10.1016/0370-2693(83)90645-7
  %%CITATION = doi:10.1016/0370-2693(83)90645-7;%%
  %152 citations counted in INSPIRE as of 11 Mar 2017
%\cite{Batalin:1984jr}
%\bibitem{Batalin:1984jr} 
%  I.~A.~Batalin and G.~A.~Vilkovisky,
%  ``Quantization of Gauge Theories with Linearly Dependent Generators,''
  Phys.\ Rev.\ D {\bf 28}, 2567 (1983)
  Erratum: [Phys.\ Rev.\ D {\bf 30}, 508 (1984)].
%  doi:10.1103/PhysRevD.28.2567, 10.1103/PhysRevD.30.508
  %%CITATION = doi:10.1103/PhysRevD.28.2567, 10.1103/PhysRevD.30.508;%%
  %1027 citations counted in INSPIRE as of 25 Mar 2017

\bibitem{BV3}
%\cite{Batalin:1985qj}
%\bibitem{Batalin:1985qj} 
  I.~A.~Batalin and G.~A.~Vilkovisky,
%  ``Existence Theorem for Gauge Algebra,''
  J.\ Math.\ Phys.\  {\bf 26}, 172 (1985).
%  doi:10.1063/1.526780
  %%CITATION = doi:10.1063/1.526780;%%
  %274 citations counted in INSPIRE as of 11 Mar 2017

%\cite{Anselmi:2008bq}
\bibitem{Anselmi:2008bq} 
  D.~Anselmi,
%  ``Weighted power counting and Lorentz violating gauge
%  theories. I. General properties,'' 
  Annals Phys.\  {\bf 324}, 874 (2009)
%  doi:10.1016/j.aop.2008.12.005
  [arXiv:0808.3470 [hep-th]];
  %%CITATION = doi:10.1016/j.aop.2008.12.005;%%
  %79 citations counted in INSPIRE as of 11 Mar 2017
%\cite{Anselmi:2008bs}
%\bibitem{Anselmi:2008bs} 
%  D.~Anselmi,
%  ``Weighted power counting and Lorentz violating gauge
%  theories. II. Classification,'' 
  Annals Phys.\  {\bf 324}, 1058 (2009)
%  doi:10.1016/j.aop.2008.12.007
  [arXiv:0808.3474 [hep-th]].
  %%CITATION = doi:10.1016/j.aop.2008.12.007;%%
  %74 citations counted in INSPIRE as of 11 Mar 2017

%\cite{Horava:2008jf}
\bibitem{Horava:2008jf} 
  P.~Horava,
%  ``Quantum Criticality and Yang-Mills Gauge Theory,''
  Phys.\ Lett.\ B {\bf 694}, 172 (2011)
%  doi:10.1016/j.physletb.2010.09.055
  [arXiv:0811.2217 [hep-th]].
  %%CITATION = doi:10.1016/j.physletb.2010.09.055;%%
  %160 citations counted in INSPIRE as of 25 Mar 2017

\bibitem{WessBagger}
J.~Wess, J.~Bagger, {\it Supersymmetry and Supergravity,}
Princeton Univ. Press (1983).

%\cite{Henneaux:1990rx}
\bibitem{Henneaux:1990rx} 
  M.~Henneaux,
  %``Space-time Locality of the {BRST} Formalism,''
  Commun.\ Math.\ Phys.\  {\bf 140}, 1 (1991).
%  doi:10.1007/BF02099287
  %%CITATION = doi:10.1007/BF02099287;%%
  %101 citations counted in INSPIRE as of 17 Apr 2017

\bibitem{Collins}
J.C.~Collins, {\it Renormalization,} Cambridge University Press (1984).

%\cite{Anselmi:2007ri}
\bibitem{Anselmi:2007ri} 
  D.~Anselmi and M.~Halat,
%  ``Renormalization of Lorentz violating theories,''
  Phys.\ Rev.\ D {\bf 76}, 125011 (2007)
%  doi:10.1103/PhysRevD.76.125011
  [arXiv:0707.2480 [hep-th]].
  %%CITATION = doi:10.1103/PhysRevD.76.125011;%%
  %99 citations counted in INSPIRE as of 12 Mar 2017

%\cite{Stelle:1977ry}
\bibitem{Stelle:1977ry} 
  K.~S.~Stelle,
%  ``Classical Gravity with Higher Derivatives,''
  Gen.\ Rel.\ Grav.\  {\bf 9}, 353 (1978).
%  doi:10.1007/BF00760427
  %%CITATION = doi:10.1007/BF00760427;%%
  %616 citations counted in INSPIRE as of 03 Apr 2017

%\cite{Vandoren:1993bw}
\bibitem{Vandoren:1993bw} 
  S.~Vandoren and A.~Van Proeyen,
  %``Simplifications in Lagrangian BV quantization exemplified by the anomalies of chiral W(3) gravity,''
  Nucl.\ Phys.\ B {\bf 411}, 257 (1994)
%  doi:10.1016/0550-3213(94)90060-4
  [hep-th/9306147].
  %%CITATION = doi:10.1016/0550-3213(94)90060-4;%%
  %37 citations counted in INSPIRE as of 17 Apr 2017

\bibitem{Polyakov}
A.~M.~Polyakov, {\it Gauge Fields and Strings,}
Harwood Academic Publishers (1987).


\end{thebibliography}
\end{document}